\documentclass[a4paper,fleqn]{aa}
\pdfoutput=1
\usepackage{natbib}
\usepackage[varg]{txfonts}
\usepackage[hidelinks]{hyperref}
\usepackage[dvipsnames]{xcolor}
\hypersetup{
    colorlinks,
    linkcolor={blue!90!black},
   citecolor={blue!90!black},
   urlcolor={blue}
}
\usepackage[T1]{fontenc}
\usepackage{ae,aecompl}
\usepackage{graphicx}
\usepackage{amsmath}
\usepackage{float}
\usepackage{amssymb}
\usepackage{multirow,bigdelim}
\usepackage{mathtools}
\usepackage{siunitx}
\usepackage{rotating}
\newcommand{\mpc}{\,h^{-1}{\rm{Mpc}}}
\newcommand{\rp}{$r_{p}$ }

\newcommand{\sqdeg}{\,{\rm{deg}}^{2}}
\newcommand{\ie}{i.e. }
\newcommand{\eg}{e.g. }
\newcommand{\pz}{photo-$z$ }
\newcommand{\vmax}{$V_{\rm{max}}$ }
\newcommand{\vmaxdc}{$V_{\rm{max,dc}}$ }
\newcommand{\zmax}{$z_{\rm{max}}$ }

\include{epsf}
\usepackage{epstopdf}
\def\apjl{{ApJ}}		% Astrophysical Journal, Letters
\def\apjs{{ApJS}}		% Astrophysical Journal, Supplement
\def\ao{{Appl.~Opt.}}		% Applied Optics
\def\aap{{A\&A}}		% Astronomy and Astrophysics
\usepackage{color}

\def \farcs{\hbox{$.\!\!^{\prime\prime}$}}

\def \sextractor{\hbox{\sc SExtractor}}
\def \sersic{\hbox{S{\'e}rsic}}

\def \galsim{\hbox{\sc GalSim}}

\defcitealias{Hoekstra17}{H17}
\defcitealias{Sheldon17}{SH17}

\newcommand{\tbf}[1]{{#1}}
\newcommand{\ttt}[1]{\texttt{#1}}

\usepackage[switch]{lineno}
\usepackage{bm}
\usepackage{soul}
\usepackage{comment}
\numberwithin{equation}{section}

\usepackage{etoolbox}

\begin{document}

\title{The PAU Survey: Intrinsic alignments and clustering of narrow-band photometric galaxies}
\titlerunning{PAUS: IA and clustering}

\author{
Harry Johnston\inst{1,2}\thanks{hj@star.ucl.ac.uk},
Benjamin Joachimi\inst{1},
Peder Norberg\inst{3,4},
Henk Hoekstra\inst{5},
Martin Eriksen\inst{6},
Maria Cristina Fortuna\inst{5},
Giorgio Manzoni\inst{4},
Santiago Serrano\inst{7,8},
Malgorzata Siudek\inst{6,9},
Luca Tortorelli\inst{10},
Jacobo Asorey\inst{11},
Laura Cabayol\inst{6},
Jorge Carretero\inst{6},
Ricard Casas\inst{7,8},
Francisco Castander\inst{7,8},
Martin Crocce\inst{7,8},
Enrique Fernandez\inst{6},
Juan Garc\'{i}a-Bellido\inst{12},
Enrique Gaztanaga\inst{7,8},
Hendrik Hildebrandt\inst{13},
Ramon Miquel\inst{6,14},
David Navarro-Girones\inst{7},
Cristobal Padilla\inst{6},
Eusebio Sanchez\inst{11},
Ignacio Sevilla-Noarbe\inst{11},
Pau Tallada-Cresp\'{i}\inst{11,15}
}

\authorrunning{H. Johnston et. al. }

\institute{
Department of Physics and Astronomy, University College London, Gower Street, London WC1E 6BT, UK\and
Institute for Theoretical Physics, Utrecht University, Princetonplein 5, 3584 CE Utrecht, The Netherlands\and
Institute for Computational Cosmology, Department of Physics, Durham University, South Road, Durham DH1 3LE, UK\and
Centre for Extragalactic Astronomy, Department of Physics, Durham University, South Road, Durham DH1 3LE, UK\and
Leiden Observatory, Leiden University, PO Box 9513, Leiden, NL-2300 RA, the Netherlands\and
Institut de F\'{i}sica  d’Altes Energies (IFAE), The Barcelona Institute of Science and Technology, Campus Univ. A. de Barcelona, 08193 Bellaterra (Barcelona), Spain\and
Institute of Space Sciences (ICE, CSIC), Campus UAB, Carrer de Can Magrans, s/n, 08193 Barcelona, Spain\and
Institut d'Estudis Espacials de Catalunya (IEEC), E-08034 Barcelona, Spain\and
National Centre for Nuclear Research, ul. Hoza 69, 00-681 Warsaw, Poland\and
Institute for Particle Physics and Astrophysics, ETH Zürich, Wolfgang-Pauli-Str. 27, 8093 Zürich, Switzerland\and
CIEMAT, Centro de Investigaciones Energéticas, Medioambientales y Tecnológicas, Avda. Complutenes 40, 28040 Madrid, Spain\and
Instituto de F\'isica Te\'orica IFT-UAM/CSIC, Universidad Aut\'onoma de Madrid, 28049 Madrid, Spain\and
Ruhr-University Bochum, Astronomical Institute, German Centre for Cosmological Lensing, Universit\"{a}tsstr. 150, 44801 Bochum, Germany\and
Instituci\'o Catalana de Recerca i Estudis Avan\c{c}ats (ICREA), 08010 Barcelona, Spain\and
Port d'Informaci\'{o} Cient\'{i}fica (PIC), Campus UAB, C. Albareda s/n, 08193 Bellaterra (Barcelona), Spain
}
% These dates will be filled out by the publisher
\date{Accepted XXX. Received YYY; in original form ZZZ}

\label{firstpage}
\makeatletter
\renewcommand*\aa@pageof{, page \thepage{} of \pageref*{LastPage}}
\makeatother

\abstract{
We present the first measurements of the projected clustering and intrinsic alignments (IA) of galaxies observed by the Physics of the Accelerating Universe Survey (PAUS). With photometry in 40 narrow optical passbands ($4500\AA-8500\AA$), the quality of photometric redshift estimation is $\sigma_{z} \sim 0.01(1 + z)$ for galaxies in the $19\,\rm{deg}^{2}$ Canada-France-Hawaii Telescope Legacy Survey (CFHTLS) W3 field, allowing us to measure the projected 3D clustering and IA for flux-limited, faint galaxies ($i < 22.5$) out to $z\sim0.8$. To measure two-point statistics, we developed, and tested with mock photometric redshift samples, `cloned' random galaxy catalogues which can reproduce data selection functions in 3D and account for photometric redshift errors. In our fiducial colour-split analysis, we made robust null detections of IA for blue galaxies and tentative detections of radial alignments for red galaxies \tbf{($\sim1-3\sigma$)}, over scales of $0.1-18\,h^{-1}\rm{Mpc}$. The galaxy clustering correlation functions in the PAUS samples are comparable to their counterparts in a spectroscopic population from the Galaxy and Mass Assembly survey, modulo the impact of photometric redshift uncertainty which tends to flatten the blue galaxy correlation function, whilst steepening that of red galaxies. We investigate the sensitivity of our correlation function measurements to choices in the random catalogue creation and the galaxy pair-binning along the line of sight, in preparation for an optimised analysis over the full PAUS area.
}

\keywords{cosmology: observations, large-scale structure of Universe}
\maketitle

\section{Introduction}

The estimation of accurate and precise galaxy redshifts over large samples is one of the major challenges in cosmology today; studies of the evolution of large-scale structure and the expansion rate of the Universe require precise knowledge of distances, which can be difficult to obtain. High-resolution spectroscopy remains an expensive technique, which is unsuited to the large volumes explored by modern wide-field galaxy surveys. Photometric redshift (photo-$z$) estimation is thus an active field of development, with various foci directed towards template-fitting (with Bayesian methods, \eg BPZ; \citealt{Benitez2000}, or maximum-likelihood fitting, \eg HyperZ; \citealt{Bolzonella2000}), empirical machine learning (\eg Directional Neighbourhood Fitting with $k$-nearest neighbours; \citealt{DeVicente2016}, or combining neural networks, decision trees, and $k$-nearest neighbours in \eg ANNz2; \citealt{Sadeh2015}), and combinations thereof (\eg training and template-fitting with {\sc{LePhare}}; \citealt{Arnouts1999}; \citealt{Ilbert2006}) -- for a recent review of \pz methods, see \cite{Salvato2019}.

The Physics of the Accelerating Universe Survey (PAUS; \citealt{Benitez2009}) tackles the \pz challenge with 40 narrow-band (NB) photometric filters, each $130 \AA$ in width, with centres in steps of $100 \AA$ from $4500 \AA$ to $8500 \AA$. In combination with existing broad-band photometry, the intermediate-resolution spectra from PAUS yield up to order-of-magnitude improvements in the precision of photo-$z$, as compared with broad-band-only estimates \citep[\eg][]{Hildebrandt2012,Marti2014,Hoyle2018,Alarcon2019}.

PAUS allows us to explore hitherto uncharted astrophysical environments, namely the weakly non-linear regime of $10-20 \mpc$ over a broad redshift epoch. PAUS straddles the boundary between (i) spectroscopic surveys, with long exposures and thus accurate redshifts, over smaller areas and volumes, and (ii) broad-band photometric surveys, having short exposures and poorer-quality redshifts, but over larger areas and volumes. PAUS thus allows for a deep, dense sampling, over an intermediate area, with unprecedented \pz precision. As such, these data offer unique snapshots of many phenomena as galaxy and small-scale processes (\eg various feedback mechanisms, non-linear density evolution) start to prompt departures from the linear regime. \tbf{Early analyses carried out for the PAU survey have included: simulations and mock catalogue generation \citep{Stothert2018}; machine learning approaches to star-galaxy classification and sky-background estimation \citep{Cabayol2019a,Cabayol2019b}; forward-modelling for narrow-band imaging \citep{Tortorelli2018}; improved photometric redshift calibration \citep{Alarcon2020,Eriksen2020}; and forecasting for Ly-$\alpha$ intensity mapping \citep{Renard2020}, among others.}

Primary science cases for PAUS include redshift-space distortions \citep[RSD;][]{Kaiser1987}, galaxy intrinsic alignments \citep[IA;][]{Heavens2000,Croft2000, Catelan2001,Hirata2004a} and galaxy clustering \citep[\eg][]{Zehavi2002}, with secondary cases for, for example, magnification \citep{Schmidt2012}. This paper focuses on the production of tailored random galaxy catalogues for PAUS, and presents initial measurements of the projected 3D intrinsic alignments and clustering of PAUS galaxies.

We developed random galaxy catalogues following the formalism laid down by \cite{Cole2011} and \cite{Farrow2015} in order to reproduce the radial selection function of the data, without any clustering along the line-of-sight. With these, and the photo-$z$ precision offered by PAUS, we are able to extend measurements of projected galaxy clustering and IA into a new regime of intrinsically faint objects up to intermediate redshifts of $z\lesssim1$, with a particular increase in statistical power for faint, blue galaxies. Our work is complementary to other direct studies of galaxy intrinsic alignments: for example, \cite{Mandelbaum2011} made null detections of alignments for bright emission-line galaxies around $z\sim0.6$ in the WiggleZ survey \citep{Drinkwater2010}; \cite{Tonegawa2017} did the same for faint, high-redshift ($z\sim1.4$), star-forming galaxies in the FastSound survey \citep{Tonegawa2015}; as did \cite{Johnston2019} when considering blue galaxies at lower redshifts in the Galaxy and Mass Assembly (GAMA) survey -- evidence continues to mount for negligible alignments in blue, spiral galaxies, going against theoretical predictions for strong alignments around the peaks of the matter distribution, which are brought on by tidal torquing mechanisms \citep[see][for a review]{Schaefer2008}. With the depth and high number density of PAUS, we push to smaller comoving galaxy-pair separations \citep{Rodriguez2020} where any signal ought to be strongest, and seek to add another data-point to the picture of blue galaxy alignments in flux-limited samples.

PAUS also presents an opportunity to attempt an extension into the faint regime of the luminosity-scaling of bright red galaxy IA found by some analyses \citep{Joachimi2011,Singh2015}. \cite{Johnston2019} found no evidence for such a scaling; however, they acknowledged complications due to satellite galaxy fractions -- the comprehensive work of \cite{Fortuna2020} modelled central and satellite, red and blue galaxy alignment contributions via the IA halo model \citep[based on][]{Schneider2010}, finding that whilst bright red galaxies might exhibit this luminosity scaling, the faint end is relatively unconstrained. With a similar redshift baseline to the luminous red galaxies (LRGs) studied by \cite{Joachimi2011} and \cite{Singh2015}, PAUS is ideally suited to assess any luminosity-dependence of IA for these fainter red galaxies, should it exist.

This paper is structured as follows; Sec. \ref{pau:sec:data} describes our galaxy data from the PAUS and GAMA surveys. In Sec. \ref{pau:sec:randoms}, we detail our construction of ‘cloned’ random galaxy catalogues. We describe the methods for measuring projected statistics in Sec. \ref{pau:sec:method}, and discuss the results in Sec. \ref{pau:sec:pauresults}. We present our concluding remarks in Sec. \ref{pau:sec:discussion}. Throughout this analysis, we quote AB magnitudes unless otherwise stated, and we compute comoving coordinates/volumes assuming a flat $\Lambda$CDM universe, with $\Omega_{m}=0.25$, $h=0.7$, $\Omega_{b}=0.044$, $n_{s}=0.95$ and $\sigma_{8}=0.8$.

\section{Data}
\label{pau:sec:data}

\begin{table*}[htpb]
    \centering
    \caption{PAUS W3 ($0.1<z_{\rm{phot.}}<0.8$) and GAMA sample characteristics: mean redshifts $\langle{z}\rangle$, mean $r$-band luminosities relative to the pivot $L_{\rm{piv}}=L(M_{r}=-22)$, the number of galaxies used as tracers of the intrinsic shear field $N_{\rm{shapes}}$, and the number of density field tracer galaxies $N_{\rm{positions}}$ -- we do not apply colour-selections to positional tracers for our intrinsic alignment correlations, using the full sample to trace the density field (see Sec. \ref{pau:sec:alignments_method}). Samples labelled `Qz$_{50}$' are defined from the best 50\% of photo-$z$ in all of PAUS W3 (\ie selected on \pz quality prior to the colour-cut). GAMA galaxies are those analysed by \cite{Johnston2019}, who limited the blue density sample objects to $M_{r}\leq-18.9$, causing the blue sample to have more shapes than density tracers.}
    
    \def\arraystretch{1.3}
    \begin{tabular}{lcccc}
    \hline
    Sample & $\langle{z}\rangle$ & $\langle{L/L_{\rm{piv}}}\rangle$ & $N_{\rm{shapes}}$ & $N_{\rm{positions}}$ \\
    \hline
    \hline
PAUS W3 red & 0.48 & 0.48 & 43725 & 43824 \\
PAUS W3 red (Qz$_{50}$) & 0.47 & 0.62 & 27427 & 27460 \\
PAUS W3 blue & 0.47 & 0.22 & 145100 & 145514 \\
PAUS W3 blue (Qz$_{50}$) & 0.44 & 0.24 & 66991 & 67209 \\
GAMA red & 0.25 & 0.77 & 69920 & 78165 \\
GAMA blue & 0.24 & 0.52 & 93156 & 89064 \\
    \hline
    \end{tabular}
    \label{pau:tab:samples}
\end{table*}

Here we provide a brief overview of the Physics of the Accelerating Universe Survey \citep[PAUS;][]{Benitez2009}, and point readers to \cite{Eriksen2019}, and references therein, for further details.

\subsection{PAU Survey}
\label{pau:sec:paus_data}

PAUS is conducted at the William Herschel Telescope (WHT), at the Observatorio del Roque de los Muchachos on La Palma, with the purpose-built PAUCam \citep{Padilla2019} instrument -- the camera sports 40 narrow-band and 6 broad-band filters on interchangeable trays, with 8 narrow-bands per tray. Each pointing is observed with between 3 and 5 dithers per tray, and exposure times for each tray are adjusted to yield as close as possible to uniform signal-to-noise ratio (S/N) as a function of wavelength -- in practice, with 8 NBs per tray, total uniformity cannot be achieved, and the S/N grows from the near-UV (limited by readout noise) to the near-IR (sky-limited).

\begin{figure*}
    \centering
    \includegraphics[width=0.8\textwidth]{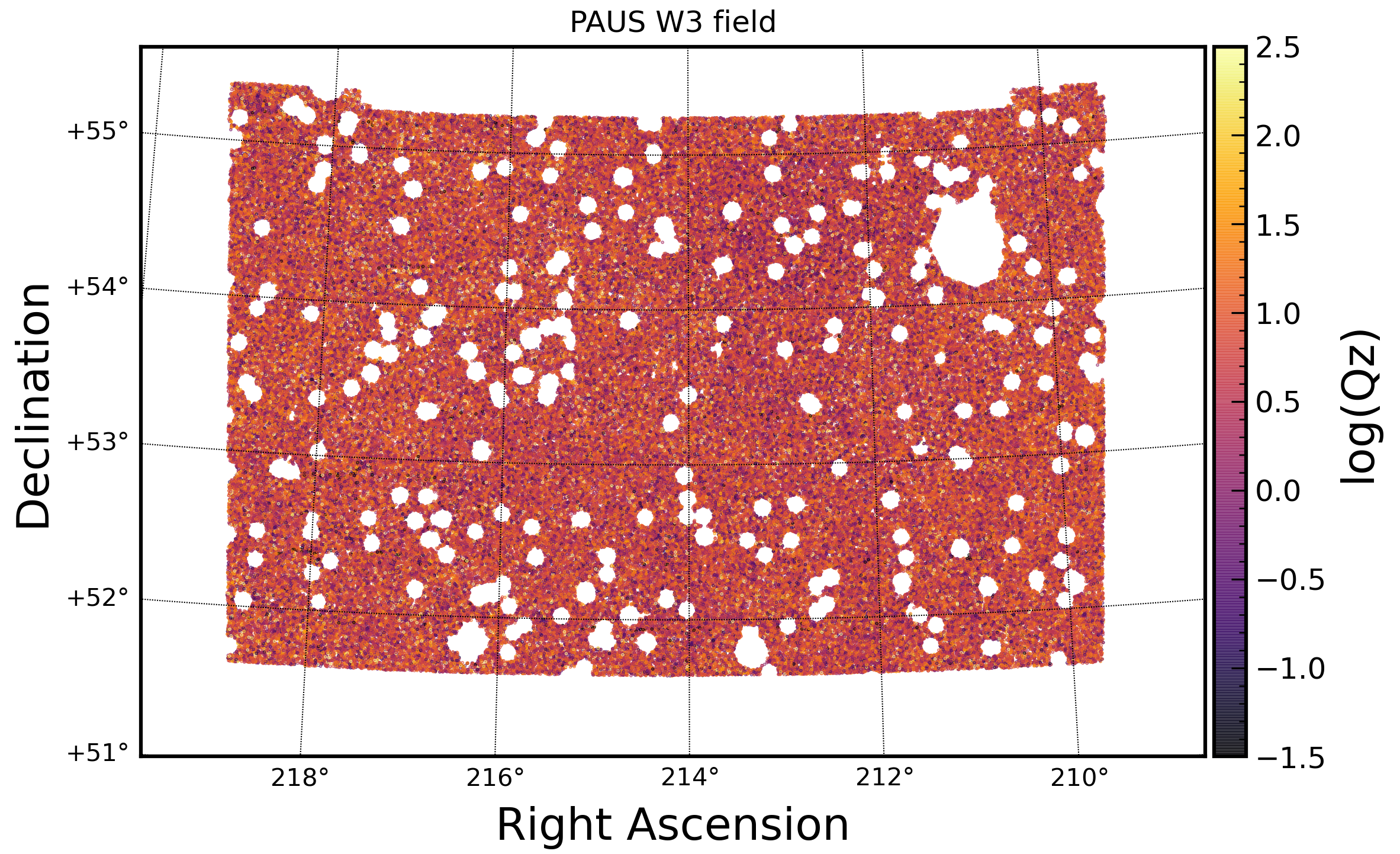}
    \caption[PAUS: W3 field footprint, coloured by photo-$z$ quality]{\tbf{Distribution of galaxies in the PAUS W3 field, with points coloured according to the log of each galaxy's \ttt{Qz} value -- this quantity describes the quality of photometric redshift estimation (low \ttt{Qz} = high photo-$z$ quality) through a combination of the goodness-of-fit of SED templates to narrow-band photometry, and the shape of the resulting redshift probability density function; see Secs. \ref{pau:sec:kcorrections} \& \ref{pau:sec:photozquality}. Circular holes in the footprint correspond to masks around foreground stars, with the largest gap (top-right) covering the Pinwheel Galaxy (Messier 101).}}
    \label{pau:fig:W3footprint}
\end{figure*}

This analysis uses PAUS data taken in the W3 field, targeting galaxies detected by the Canada-France-Hawaii Telescope Legacy Survey \citep[CFHTLS;][]{Cuillandre2006}. The current PAUS coverage of W3 is $\sim19\sqdeg$ -- the completed PAU Survey aims to cover $\sim100\sqdeg$ over several non-contiguous fields -- and we retain {263\,227} galaxies for analysis after imposing the flux-limit $i<22.5$, and rejecting stars \citep[see][for CFHTLenS \ttt{star\_flag} details]{Erben2013} and bright sources with $i<18$, and sources for which we were missing five or more narrow-bands. \tbf{The PAUS W3 field footprint is displayed in Fig. \ref{pau:fig:W3footprint}, where galaxy points are coloured according to our photo-$z$ quality marker \ttt{Qz} (lower values = better photo-$z$; see Secs. \ref{pau:sec:kcorrections} \& \ref{pau:sec:photozquality}) -- one sees that there are no obvious trends in photo-$z$ quality with respect to positions on-sky. For our correlation function analysis, we reject objects for which shape estimation failed completely}\footnote{\tbf{These were found to be randomly distributed across the footprint, but with significantly poorer photo-$z$; these shape/redshift failures will be investigated in an upcoming, full-area correlation function analysis.}}, and further restrict our PAUS samples to {189\,338} galaxies within the \pz range $0.1<z_{\rm{phot.}}<0.8$ (for reasons we discuss in Sec. \ref{pau:sec:photozquality}).

\subsection{{\sc{bcnz2}} \& $k$-corrections}
\label{pau:sec:kcorrections}

The {\textsc{bcnz2}} photometric redshift algorithm -- presented in detail by \cite{Eriksen2019} -- was designed to tackle the challenge of photo-$z$ estimation with 40 optical narrow-bands, supplementary broad-bands (from CFHTLS), and the flexible utilisation of galaxy emission lines -- the latter point in particular is important, since many of the high-redshift, blue galaxies targeted by PAUS lack large spectral breaks. As \cite{Eriksen2019} show, strong emission lines from these galaxies can be leveraged to achieve powerfully precise photo-$z$ -- reaching accuracies of $\sigma_{z}=0.0037(1+z)$ for the best 50\% of photo-$z$ derived for the COSMOS field. Descriptors for the quality of photo-$z$ estimates are explored by \cite{Eriksen2019} -- we make use of the \ttt{Qz} parameter (their Sec. 5.5) when assessing photo-$z$ in W3, and the impacts of quality-cuts.

\begin{figure}
    \centering
    \includegraphics[width=\columnwidth]{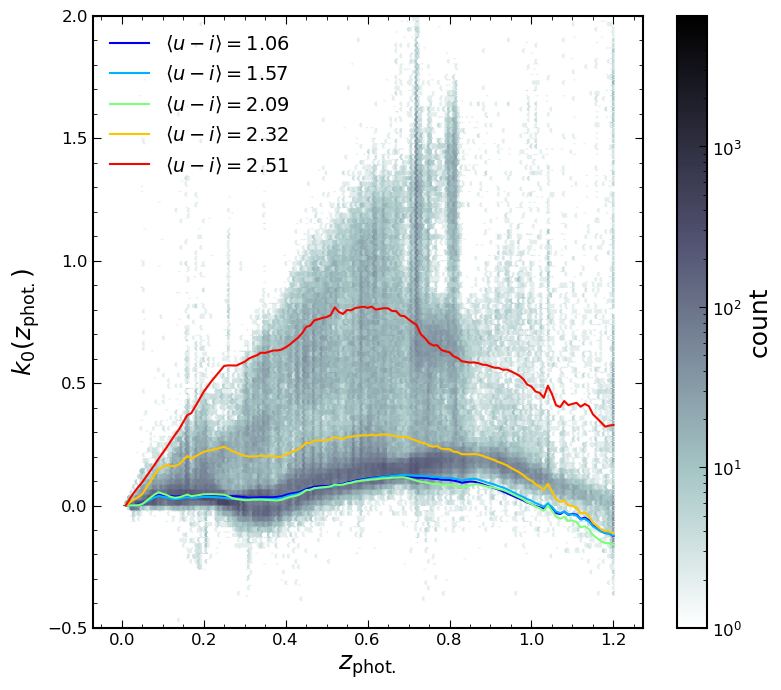}
    \caption[PAUS: $i$-band $k$-corrections to $z=0$ as a function of redshift and rest-frame colour]{Hexagonally-binned 2D histogram of PAUS W3 galaxies' $i$-band $k$-corrections from redshift $z$ (x-axis) to $z=0$ -- we assembled these by redshifting each galaxy's best-fit SED model over the entire $z$-range, and taking ratios (Eq. \ref{pau:eq:paus_uniq_kcorrection}) of fluxes to the $z=0$ flux. Cells are coloured by the count of galaxies resident in each cell. Solid coloured lines give the running-medians of $k_{0}(z)$ for five rest-frame colour bins, for which the average colours $\langle{}u-i\rangle$ are given in the legend (absolute magnitudes estimated with {\sc{LePhare}}).}
    \label{pau:fig:paus_kcorrections}
\end{figure}

\begin{figure*}
    \centering
    \includegraphics[width=\linewidth]{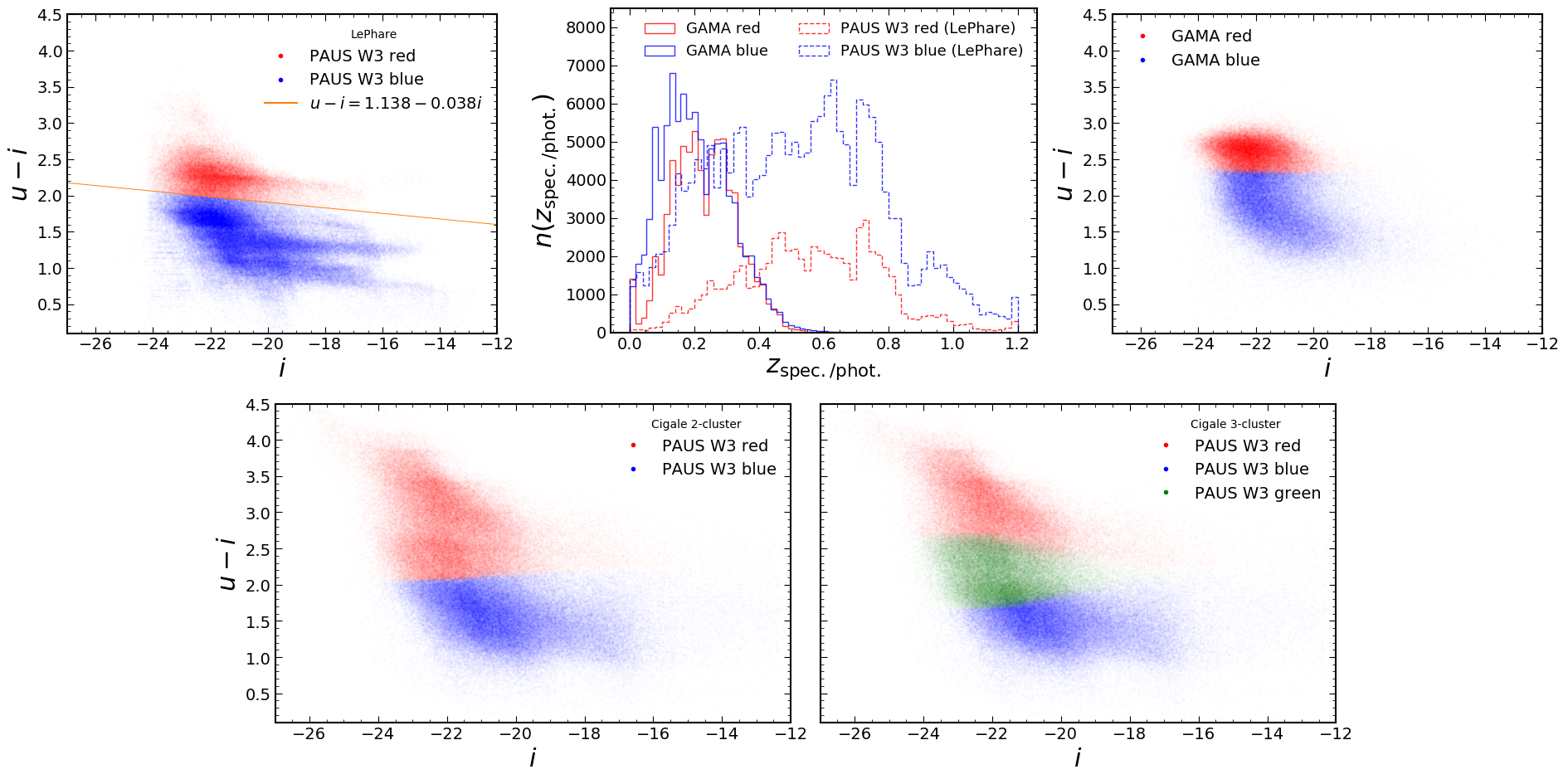}
    \caption[PAUS and GAMA: rest-frame $u-i$ colour vs. $i$-magnitude diagrams and photometric/spectroscopic redshift distributions for red/blue galaxies]{\emph{Top-left:} Absolute rest-frame colour $u-i$ vs. $i$ magnitude for PAUS galaxies in the W3 area. These rest-frame magnitudes were derived for CFHTLenS \citep{Erben2013} using the {\textsc{LePhare}} software package (\citealt{Arnouts1999}; \citealt{Ilbert2006}). We define the red sequence as those galaxies for whom $u-i>1.138-0.038i$, as indicated by the orange line. \emph{Top-middle:} Spectroscopic redshift distribution of galaxies in GAMA, and photo-$z$ distribution of PAUS W3, coloured according to the cuts in the top-left and top-right panels -- we note that PAUS samples are restricted to $0.1<z_{\rm{phot.}}<0.8$ for our correlation function analysis. \emph{Top-right:} Absolute rest-frame colour $u-i$ vs. $i$ magnitude for GAMA equatorial galaxies from the final GAMA data-set \citep{Liske2015}. We followed \cite{Johnston2019} in setting the red/blue boundary at rest-frame $g-r=0.66$, and plot the galaxies on the $u-i$ vs. $i$ plane for comparison with PAUS W3. \emph{Bottom:} {\sc{Cigale}}-estimated absolute rest-frame $u-i$ vs. $i$ for PAUS W3, with 2-cluster (\emph{left}) and 3-cluster (\emph{right}) galaxy type classification.}
    \label{pau:fig:PAUGAMAzCMD}
\end{figure*}

\begin{figure}
    \centering
    \includegraphics[width=\columnwidth]{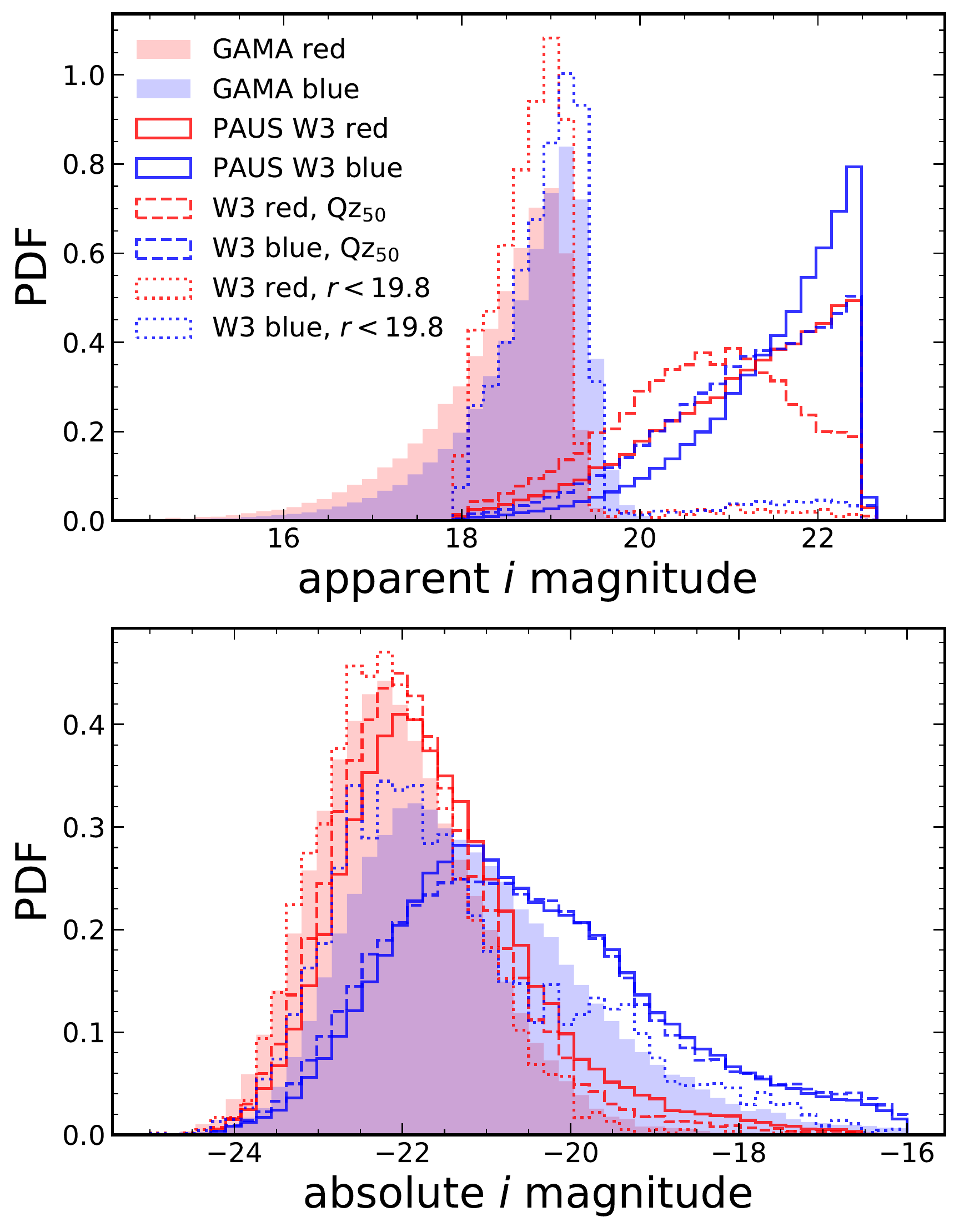}
    \caption[PAUS and GAMA: apparent and absolute rest-frame $i$-band magnitude distributions for red/blue galaxies]{Apparent (observed) and absolute $i$-band magnitudes of galaxies in PAUS (solid lines; limited to $0.1<z_{\rm{phot}}<0.8$ and apparent $i>18$) and GAMA (shading), with colours reflecting the ({\sc{LePhare}}) red/blue selections displayed in Fig. \ref{pau:fig:PAUGAMAzCMD} (top-left panel). We also display the magnitude distributions for 2 subsets of PAUS: (i) the best 50\% selected on \pz quality parameter \ttt{Qz} (dashed), and (ii) with apparent $r<19.8$ (dotted) -- approximately matching the GAMA flux-limit. Each histogram is individually normalised to unit area under the curve. Whilst the red galaxies in PAUS/GAMA are fairly similarly distributed in absolute $i$-band magnitude, PAUS exhibits a higher fraction of intrinsically faint blue galaxies. PAUS absolute magnitudes shown here are those computed using {\sc{LePhare}}.
    }
    \label{pau:fig:gama_pau_magnitudes}
\end{figure}

The method of {\textsc{bcnz2}} involves fitting to the narrow-band flux data with linear combinations of template galaxy SEDs \tbf{(based upon \citealt{Bruzual2003}, and described in \citealt{Eriksen2019}; Sec 4.4)} -- as a by-product of this procedure, we gain a best-fitting SED model for each object, which we can redshift arbitrarily. From these models we are able to easily compute unique $k$-corrections per object, for a given band, via

\begin{equation}
    k_{z}(z_{\rm{obs.}}) = -2.5 \,{\rm{log}}\left(\frac{f_{z_{\rm{obs.}}}}{f_{z}}\right) \quad ,
    \label{pau:eq:paus_uniq_kcorrection}
\end{equation}
where $f_{z}$ is the flux transmitted to the observer, in that waveband, by an object at redshift $z$; thus $f_{z_{\rm{obs.}}}$ is the observed flux of the object in that band. The $k$-correction $k_{z}(z_{\rm{obs.}})$ then modifies the flux of a galaxy at redshift $z_{\rm{obs.}}$ to look as though it were at redshift $z$; $k$-corrections are thus necessary to infer the maximum redshift $z_{\rm{max}}$ at which a galaxy can be observed by a flux-limited survey. Fig. \ref{pau:fig:paus_kcorrections} displays $k_{0}(z)$ for PAUS W3 galaxies -- these are the $k$-corrections from a given redshift $z$ to $z=0$. Given that $k$-corrections are known to correlate strongly with galaxy type (via the archetypal forms of SEDs), we also defined a set of $k$-corrections by binning galaxies according to their rest-frame absolute $u-i$ colour (estimated using {\sc{LePhare}} -- see below), before taking running medians of $k_{0}(z)$ for each colour-bin \citep{McNaught-Roberts2014} -- these median $k$-corrections are displayed as solid lines in Fig. \ref{pau:fig:paus_kcorrections}. We found however, that the colour-median and the unique (per-galaxy) corrections yielded negligibly different estimates for the redshifts $z_{\rm{max}}$ at which galaxies (of fixed magnitude) cross the survey flux-limit -- to be discussed in Sec. \ref{pau:sec:randoms}.

\subsection{Rest-frame magnitudes \& colours}
\label{pau:sec:restframe_mags_colours}

When quoting rest-frame magnitudes, or estimating PAUS W3 galaxy colours, we make use of two independent determinations of these quantities: (i) those derived for CFHTLenS galaxies \citep[see][]{Erben2013} using the {\textsc{LePhare}} \citep{Arnouts1999,Ilbert2006} package, and (ii) those we have determined for PAUS, using the {\sc{Cigale}} \citep{Noll2009,Boquien2019} software package. The former quantities were derived with low-resolution photometry and are consequently more noisy/prone to biases and degeneracies in redshift-colour space -- such as those clearly visible in Fig. \ref{pau:fig:PAUGAMAzCMD} (top-left panel). The rest-frame colours we have derived with {\sc{Cigale}} make use of the full complement of PAUS photometry, with 40 narrow-bands and 6 CFHTLS broad-bands, yielding smoother colour-magnitude distributions (bottom-panels). We draw a line $u-i=1.138-0.038i$ on the PAUS {\sc{LePhare}} colour-magnitude plane, above which we classify galaxies as `red' (early-type) with `blue' (late-type) galaxies below -- this boundary is fairly arbitrary, chosen only to separate a dense red sequence from the more diffuse blue cloud, each visible in the top-left panel of Fig. \ref{pau:fig:PAUGAMAzCMD}. For {\sc{Cigale}} colours, we have separated galaxy types using 2- or 3-cluster classifications \citep[as described in][]{Siudek2018,Siudek2018a} in the multidimensional rest-frame colour-magnitude space: \{$i$, $g-i$, $r-z$, $g-r$, $u-g$\}, resulting in definitions for red-sequence and blue-cloud galaxy samples with some interlopers (2-cluster), or red/blue samples with theoretically greater purity, having excluded the green valley (3-cluster). We shall measure and compare correlations in PAUS with each of these three different colour classifications.

Fig. \ref{pau:fig:PAUGAMAzCMD} also displays the photometric redshift distributions of our red and blue PAUS galaxy samples (top-middle panel), along with spectroscopic redshifts for red and blue galaxies from the GAMA survey \citep[][see Sec. \ref{pau:sec:paus_gama}]{Driver2009}, where GAMA samples are split according to a boundary at rest-frame $g-r=0.66$ \citep[following][]{Johnston2019}. Fig. \ref{pau:fig:gama_pau_magnitudes} then compares the distributions of apparent and absolute $i$-band magnitudes between our PAUS (solid-step histograms) and GAMA (shaded histograms) samples, including when selecting on PAUS for \pz quality (best 50\% via \ttt{Qz}; dashed histograms), or to approximately match the GAMA flux-limit of $r<19.8$ (dotted histograms). From Figs. \ref{pau:fig:PAUGAMAzCMD} \& \ref{pau:fig:gama_pau_magnitudes}, one sees how PAUS is complementary to GAMA; PAUS offers insight into a different population of fainter, bluer galaxies over a long redshift baseline, making it ideal for studying galaxy correlations as functions of environment and redshift. \tbf{We neglect to conduct a detailed search for any redshift evolution of galaxy correlations in this work, as we are currently lacking the statistical power to do so with precision; future PAUS work with increased area will explore the redshift dependence of signals, along with dependencies on galaxy colour, luminosity etc.}

\subsection{Galaxy shape estimation}

To measure the shapes of the galaxies we use weighted quadrupole moments\footnote{\tbf{Other shape estimation methods, \eg model-fitting, such as in \emph{lens}fit \citep{Miller2013}, are often not optimised for measuring the shapes of apparently bright galaxies, as these form only a small fraction of a typical cosmic shear catalogue. Since our PAUS sample has a fair number of such objects, and is not so deep that moments-based methods begin to suffer, we elect to make use of moments for shape estimation in this work.}} $I_{ij}$ which are defined as
\begin{equation}
I_{ij}=\frac{1}{I_0}\int{\rm d}^2{\bm x}\,x_i\, x_j\, W({\bm x})\,f({\bm x}) \quad ,
\end{equation}
where $f({\bm x})$ is the observed \tbf{$i$-band} galaxy image (flux), $W({\bm x})$ is a suitable weight function to suppress the noise, $I_0$ is the weighted monopole moment, and $x_i,x_j$ denote the image coordinate axes. The moments are combined to form the `polarisation', which quantifies the shape
\begin{equation}
e_1=\frac{I_{11}-I_{22}}{I_{11}+I_{22}},~{\rm and~}
e_2=\frac{2I_{12}}{I_{11}+I_{22}} \quad .
\end{equation}
The resulting shapes are, however, biased; firstly, the weight function changes the quadrupole moments with respect to the unweighted case. Although this reduces the noise in the quadrupole moments, the estimate of the polarisation involves a ratio of moments that are noisy themselves. This leads to the so-called `noise bias' \citep{Kacprzak12,Viola14}. Finally, the observed image $f({\bm x})$ is convolved with the point spread function (PSF).

A wide range of algorithms has been developed to relate the observations to unbiased estimates of the gravitational lensing shear. Our objective is very similar, although we note that an unbiased shear estimate is not quite the same as an unbiased ellipticity estimate. Here we use the algorithm developed by \cite*{Kaiser1995} and \cite{LK97}, with modifications described in \cite{Hoekstra98}, to correct
the observed polarisations for both the weight function and the blurring by an (anisotropic) PSF. We refer the reader to these papers for further details. 

The estimate for the ellipticity\footnote{This is a shear estimate, strictly speaking, which we instead use as a proxy for the ellipticity $\epsilon$.} is given by
\begin{equation}
    \epsilon^{\rm KSB}_i=\frac{e_i-P_{ii}^{\rm sm} p_i}{P^\gamma} \quad ,
    \label{pau:eq:shear_estimate}
\end{equation}
where the smear polarisability $P^{\rm sm}$ quantifies the response to the smearing by the PSF, and $p_i\equiv e^{\rm *}_i/P^{\rm sm,*}_{ii}$ captures the PSF properties. The pre-seeing shear polarisability $P^\gamma$ corrects for the circularisation of the shapes by both the PSF and the weight function. 
Formally a $2\times 2$ tensor, we assume it is diagonal with both elements having the same amplitude. Both polarisabilities involve higher order moments and 
\begin{equation}
    P^\gamma\equiv P^{\rm sh}-P^{\rm sm}\left(\frac{P^{\rm sh,*}}{P^{\rm sm,*}}\right) \quad ,
    \label{eq:shear}
\end{equation}
is a combination of the shear polarisability $P^{\rm sh}$ which captures the response to a shear for weighted moments, and the smear polarisability of both the galaxy and the PSF. As shown in \cite{Hoekstra98}, the PSF moments and polarisations should be measured using the same weight function as was used for the galaxies.

In principle, one is free to choose any weight function to estimate the galaxy ellipticity, but for intrinsic alignment studies this may affect the signal: a broader weight function is more sensitive to the outskirts of a galaxy compared to a more compact kernel. As tidal processes typically affect larger galactic radii, the IA signal may then depend on the choice of filter width. This was confirmed most convincingly by \cite{Georgiou2019a}, but also see \cite{Singh16}.

To allow for a comparison with the IA signals presented in \cite{Johnston2019}, we choose the width of the weight function so that it resembles the one employed by \cite{Georgiou2019b}. The weight function for the bright galaxies studied in these papers was based on an isophotal limit: $r_{\rm iso}=\sqrt A_{\rm iso}/\pi$, where $A_{\rm iso}$ is the area above $3\times$ the background noise level as determined by \sextractor \,\,\citep{Bertin96}.
Although less susceptible to prominent bulges in well-resolved galaxies, this definition is difficult to link to the weight functions typically used in weak lensing studies. Moreover, the size depends upon the depth of the particular data set used, and thus is not uniquely defined. 

In weak lensing studies, the width of the weight function matches the size of the observed galaxy image. Although this depends somewhat on the image quality, in practice it is better defined. As a compromise, however, we increase the width of the weight function to 1.75 times the observed half-light radius of the galaxy, as we found that this roughly matches the width used by 
\cite{Georgiou2019b}.

\subsubsection{Calibration of estimated ellipticities}
\label{pau:sec:shape_calibration}

Although Eq.~\ref{eq:shear} yields decent estimates for the ellipticities of the relatively bright galaxies considered here, these estimates are still biased. Moreover, the use of a wider weight function will increase the noise bias. To account for these biases, we follow \cite{Hoekstra15} and create simulated images to determine the multiplicative bias correction as a function of observing conditions; that is to say, the seeing and galaxy properties dictating objects' size and S/N.

The setup of the image simulations is similar to the one used in \cite{Hoekstra15}, and we refer the interested reader to that paper for greater detail. The galaxy properties are drawn from a catalogue of morphological parameters that were measured from resolved $F606W$ images from the Galaxy Evolution from Morphology and SEDs survey \citep[GEMS;][]{Rix04}. These galaxies were modelled as
single \sersic\ models with {\tt galfit} \citep{Peng02}.
We assume that the morphological parameters of the galaxies do not depend on the passband, although we note that our calibration should be able account for such differences. The background noise level is matched to the average value in the CFHTLS $i$-band data. The background varies in the data, but again, our calibration uses the S/N, which naturally accounts for the variation in noise level
\citep[see][]{Hoekstra15}.

To measure the bias in the simulations we match the galaxies to the input catalogue and determine the best fit 
\begin{equation}
\epsilon^{\rm KSB}_i=(1+{\mu}_i)\epsilon^{\rm input}_i + c_i \quad ,
\label{eq:ellipticity_bias}
\end{equation}
where ${\mu}_i$ is the multiplicative bias and $c_i$ the additive bias, which is found to be zero for our axisymmetric PSF. Moreover, we find that ${\mu}_1$ and 
${\mu}_2$ agree with one another, so we only consider the average bias ${\mu}$ henceforth.

Similarly to \cite{Hoekstra15}, we assume that the multiplicative bias is predominantly determined by the angular size of the galaxy and the S/N. As a proxy for the size of the galaxy before convolution by the PSF, we define the parameter ${R}$ as

\begin{equation}
{R}=\sqrt{r^2_{\rm{h, obs}}-r^2_{\rm h,*}} \quad ,
\end{equation}
where $r_{\rm h,*}$ denotes the half-light radius of the PSF and $r_{\rm h, obs}$ that of the observed galaxy. However, to better capture the PSF dependence, we create images with seeing ranging from $0\farcs 5$ to $1\farcs 2$ and derive the correction as a function of seeing.

For a given PSF size, we determine ${\mu}$ in narrow bins of signal-to-noise ratio $\nu$ and galaxy size-proxy ${R}$. We found that the multiplicative bias ${\mu}$ can be parameterised fairly well by 
\begin{equation}
    {\mu}(\nu,{R})\,\big|_{i<22.5} = \alpha_0+\frac{\alpha_1}{R}+\frac{\alpha_2}{\nu}+\frac{\alpha_3}{\nu^2}+\alpha_4 \sqrt{\frac{R}{\nu}} \quad ,
    \label{eq:mufit}
\end{equation}
where the parameters $\alpha_j$ are determined by fitting this model to the estimates of ${\mu}$ from Eq.~\ref{eq:ellipticity_bias}. 

\begin{figure}
\centering
\leavevmode \hbox{% 
  \includegraphics[width=8.5cm]{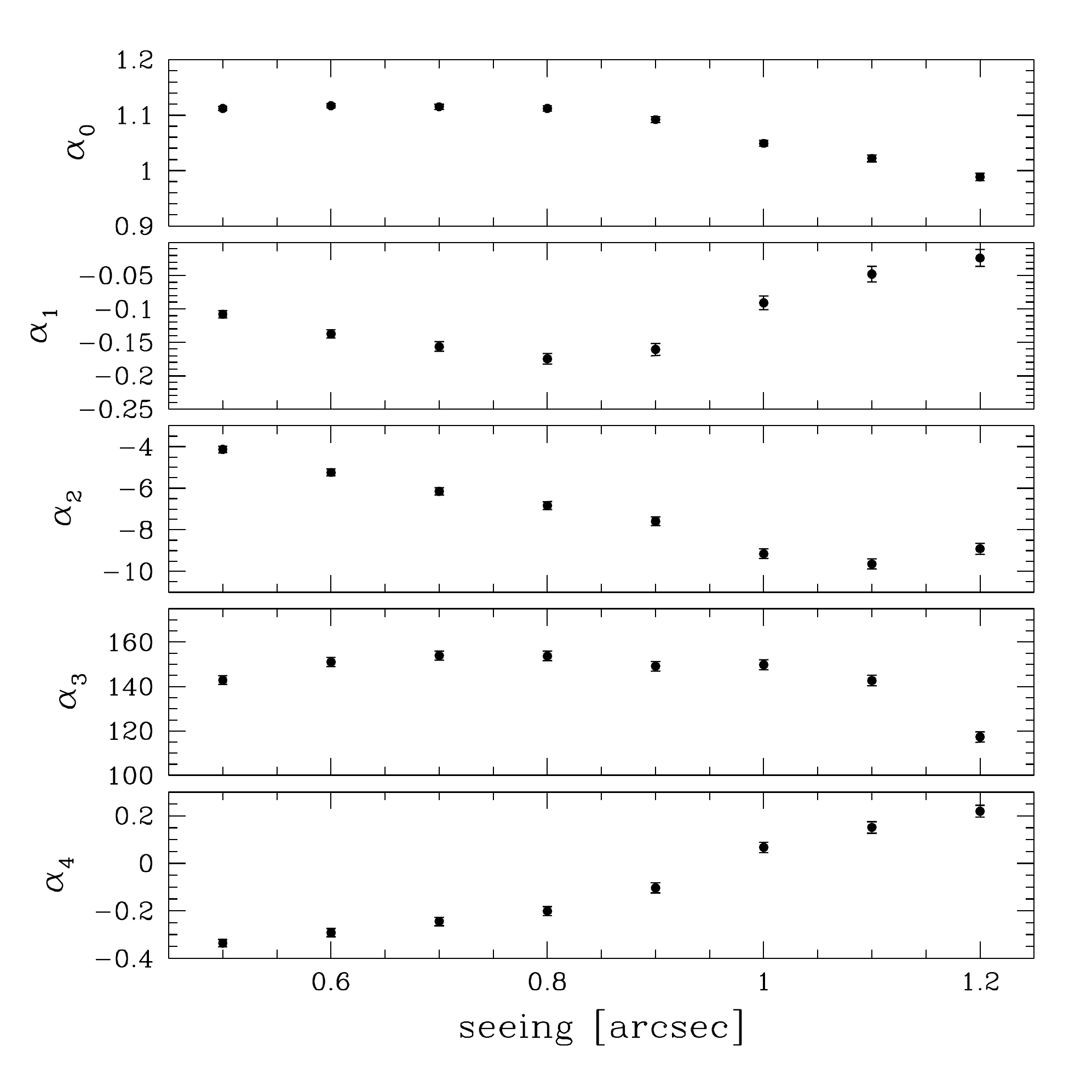}}
\caption{Best fit parameters $\alpha_j$ as a function of seeing. These parameters are used to estimate the multiplicative bias as a function of galaxy size and signal-to-noise ratio.
  \label{fig:fitpar}}
\end{figure}

As the IA measurements are limited to galaxies with $i<22.5$, we calibrate the multiplicative bias for this range in magnitude. Fig.~\ref{fig:fitpar} shows the values of the model parameters as functions of seeing. The resulting parameters minimise the bias for our sample, but we verified that the biases are also significantly reduced outside this magnitude range. Nonetheless, the residual bias does vary with magnitude, and the correction needs to be recomputed if different magnitude ranges are considered, because it changes the underlying population of galaxies \citep[see][]{Kannawadi19}.

\begin{figure}
\centering
\leavevmode \hbox{% 
  \includegraphics[width=8.5cm]{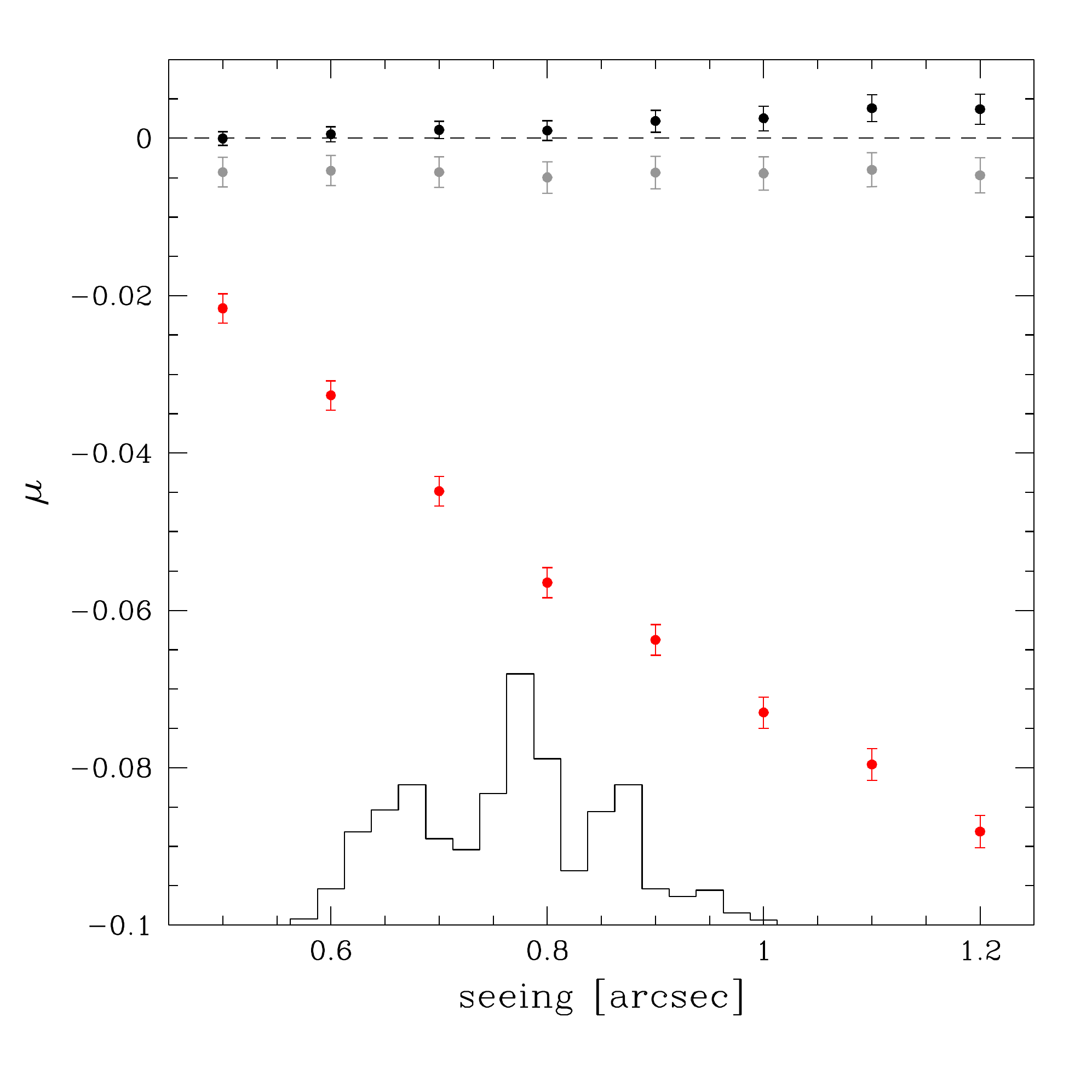}}
\caption{Multiplicative bias ${\mu}$ as a function of seeing for simulated galaxies with $20<i<22.5$. The red points correspond to the KSB estimates, which are significantly biased and depend upon image quality. The corrected biases are shown as grey (per-galaxy bias correction) and black (seeing-averaged bias correction) points. 
The histogram shows the seeing distribution for the CFHTLS W3 data.
  \label{fig:bias_seeing}}
\end{figure}

Fig.~\ref{fig:bias_seeing} shows the multiplicative bias ${\mu}$ as a function of seeing for simulated galaxies with $20<i<22.5$, overlaid with the seeing distribution from the measured PSF sizes for each of the CFHTLS Megacam chips. The red points are the \cite*{Kaiser1995} (KSB) estimates, which show a clear dependence on seeing. Without an additional correction the mean bias $\langle{\mu}_{\rm KSB}\rangle=-0.051$. 

\tbf{
Eq.~\ref{eq:mufit} and the resulting fitted parameters in Fig.~\ref{fig:fitpar} describe a model to estimate the required multiplicative ellipticity bias correction, as a function of a given object's size $R$ and signal-to-noise ratio $\nu$. One approach to calibration is to use the model to apply corrections to each simulated galaxy individually; we do this and show the residual bias (grey points) in Fig.~\ref{fig:bias_seeing}. The mean residual bias for this approach is $\langle{\mu}_{\rm cor}\rangle=-0.0045$. Alternatively, we can average the model estimates of the bias within bins of seeing, and apply these averaged corrections to the galaxies falling in each bin, thus lessening the impact of noise from individual galaxy shapes. These corrections are shown in Fig.~\ref{fig:bias_seeing} as black points, with a mean of $\langle{\mu}_{\rm cor}\rangle=0.0012$ -- slightly better than the first approach, and sufficient for our IA work here.
}

%To correct the KSB shapes we can use Eq.~\ref{eq:mufit} to estimate a bias per galaxy, or we can average the bias first and use this average bias to correct. The grey points show the results for the former approach, which yields $\langle{\mu}_{\rm cor}\rangle=-0.0045$, whereas the latter approach (black points) performs slightly better with $\langle{\mu}_{\rm cor}\rangle=0.0012$, sufficient for the IA measurements presented here.

As we aim to measure the IA signal for samples split by morphology (which roughly correlates with our colour-split) it is worth examining whether our correction can be used in this case. We measured residual bias when we split the sample at input \sersic-index $n_{\rm{in}}=2$, chosen for indicative purposes \citep[see \eg][Fig. 2]{Vakili2020}. For galaxies with $n_{\rm in}\le 2$ we find 
$\langle{\mu}_{\rm cor}\rangle=0.0048$ and for $n_{\rm in}>2$ we obtain $\langle{\mu}_{\rm cor}\rangle=-0.0114$; although the bias depends on the radial surface brightness profile, the differences are too small to affect our results. In any case, we calibrate our PAUS red and blue sample ellipticities according to the seeing-averaged multiplicative biases calculated within each colour-split sample.

%We estimate galaxy ellipticities via the `polarisation' $\epsilon=(a-b)^{2}/(a+b)^{2}$, using the method of \cite*{Kaiser1995} -- commonly referred to as `KSB'. $a$ and $b$ are the galaxy semi-major and semi-minor axes, respectively. The KSB technique extracts object ellipticities from the weighted quadrupole moments of the flux distribution across images, and derives corrections for the smearing effects of anisotropic point-spread functions (PSFs) by fitting models to the isophotes of foreground stars.

\subsection{Photo-$z$ quality}
\label{pau:sec:photozquality}

\begin{figure}
    \centering
    \includegraphics[width=\linewidth]{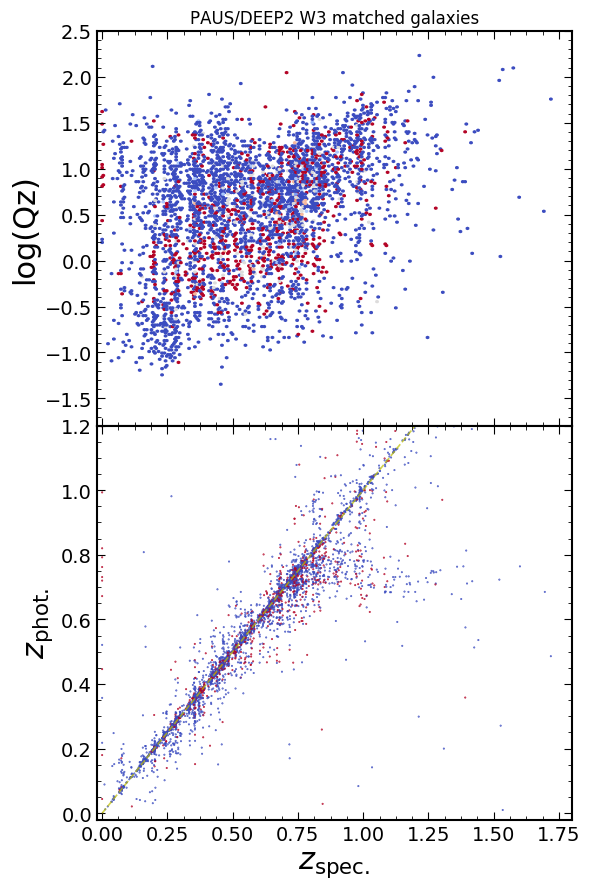}
    \caption[PAUS: photo-$z$ and photo-$z$ quality (\ttt{Qz}) as a function of spectroscopic redshift for red/blue PAUS-DEEP2 matched galaxies]{\emph{Top:} Photo-$z$ quality $\rm{log}(\ttt{Qz})$ for {$\sim$4k} PAUS-DEEP2 matched galaxies, as a function of their spectroscopically determined redshifts $z_{\rm{spec.}}$, with red and blue colours reflecting the {\sc{LePhare}} classification of the galaxies (see Fig. \ref{pau:fig:PAUGAMAzCMD}, top-left panel). A smaller value of \ttt{Qz} indicates a higher-quality photo-$z$ estimate. \emph{Bottom:} PAUS photo-$z$ estimates vs. DEEP2 spec-$z$, again coloured by galaxy classification. The 1-to-1 relation is shown in the bottom panel as a faint yellow dashed line.
    }
    \label{pau:fig:zphot_zspec_qz}
\end{figure}

\begin{figure}
    \centering
    \includegraphics[width=\linewidth]{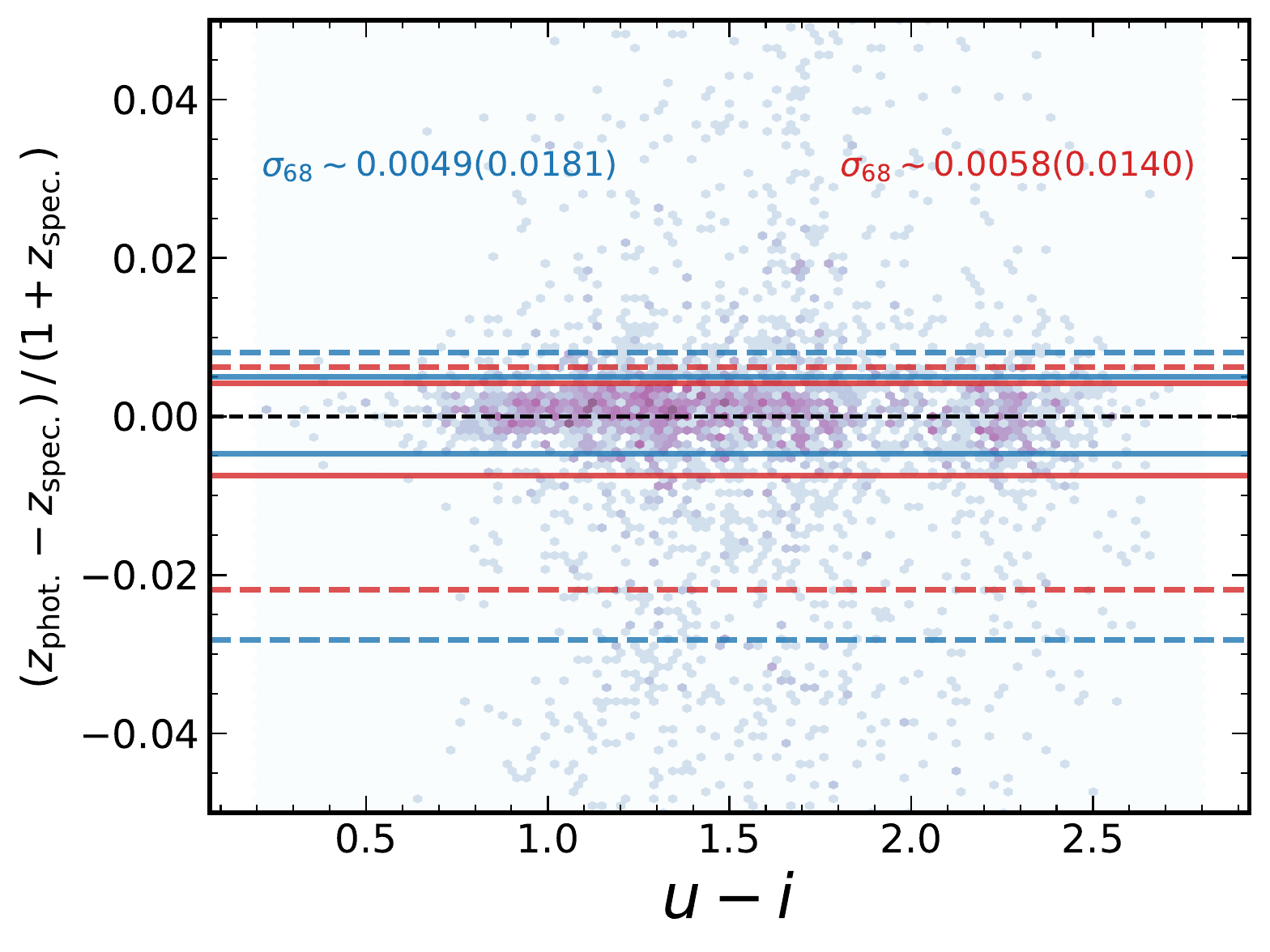}
    \caption[PAUS: normalised photo-$z$ error as a function of rest-frame colour $u-i$ for PAUS-DEEP2 matched galaxies]{Hexagonally-binned 2D histogram of PAUS-DEEP2 matched galaxies' photo-$z$ error, normalised by their spectroscopic redshifts and as a function of their {\sc{LePhare}} rest-frame colour $u-i$. Shading indicates the log-counts in each cell. For the best 50\% of \pz according to the \ttt{Qz} parameter, we display the $16^{\rm{th}},84^{\rm{th}}$ percentiles of the normalised photo-$z$ error for red (blue) galaxies as solid red (blue) horizontal lines. Half the difference between the percentiles is quoted for each as $\sigma_{68}$. Dashed lines and bracketed values of $\sigma_{68}$ are then for the total sample.}
    \label{pau:fig:zphot_colour}
\end{figure}

We investigate the quality of the photo-$z$ estimates in W3, plotting the \ttt{Qz} parameter\footnote{As \cite{Eriksen2019} describe in their Sec. 5.5, the \ttt{Qz} parameter measures \pz quality through a combination of the width of the redshift probability density function $P(z)$, the probability volume surrounding its peak, and the $\chi^{2}$ of the template-fit to the galaxy SED.} and $z_{\rm{phot.}}$ against $z_{\rm{spec.}}$ for {$\sim$4k} PAUS galaxies matched to the spectroscopic DEEP2 \citep{Newman2013} \href{http://deep.ps.uci.edu/DR4/zcatalog.html}{DR4} catalogue in Fig. \ref{pau:fig:zphot_zspec_qz}. We see that the quality of $z_{\rm{phot.}}$ drops with increasing $z_{\rm{spec.}}$ (small \ttt{Qz} = good $z_{\rm{phot.}}$), and that there is no drastic performance differential between red and blue galaxies. Fig. \ref{pau:fig:zphot_colour} looks deeper, considering the redshift error $z_{\rm{phot.}}-z_{\rm{spec.}}$, normalised by $1+z_{\rm{spec.}}$, as a function of the {\sc{LePhare}} rest-frame colour $u-i$. The $16^{\rm{th}},84^{\rm{th}}$ percentiles are shown for each colour as dashed lines, and as solid lines for the best 50\% of \pz according to the \ttt{Qz} parameter. For the \ttt{Qz}-selection, half the difference between the percentiles is quoted for each colour as $\sigma_{68}$, with the full-sample $\sigma_{68}$ given in brackets. We see that red galaxies have marginally lower quality $z_{\rm{phot.}}$ than blue galaxies when selecting on \ttt{Qz}, and that this behaviour is reversed over the full sample, with red \pz performing better. We also see that all $z_{\rm{phot.}}$ have a tendency to be under-estimated with respect to the spectroscopic redshifts.

%whereas the blue $z_{\rm{phot.}}$ errors are more symmetric about zero. The higher quality of photo-$z$ for blue galaxies is due to their relative wealth of spectral emission lines, which are well-sampled by the 40 PAUS narrow-bands, but atypical for the spectra of red galaxies with low rates of star formation.

Considering the sparsity of PAUS W3 objects beyond $z_{\rm{phot.}}\sim0.8$ (Fig. \ref{pau:fig:PAUGAMAzCMD}), and the small volume probed (by just $\sim19\sqdeg$) at $z_{\rm{phot.}}\lesssim0.1$, we restrict our measurements to PAUS galaxies in the range $0.1<{z_{\rm{phot.}}}<0.8$. We note that Fig. \ref{pau:fig:zphot_zspec_qz} suggests a drop in the quality of photo-$z$ at $z_{\rm{phot.}}\sim0.7$ -- we discuss the potential for mitigation of \pz errors with random galaxy catalogues in Sec. \ref{pau:sec:photoz_randomsimpact}.

\subsection{GAMA}
\label{pau:sec:paus_gama}

The Galaxy and Mass Assembly (GAMA; \citealt{Driver2009}; \citealt{Driver2011a}; \citealt{Baldry2018}) survey was conducted at the Anglo-Australian Telescope, and the final data were described by \cite{Liske2015}. GAMA achieved high completeness ($\gtrsim98\%$) down to $r<19.8$ over $180\sqdeg$ within the equatorial fields G9, G12, G15 ($60\sqdeg$ apiece). We compare the correlation functions we measure in PAUS with analogous measurements made in GAMA and presented in \cite{Johnston2019}, where details of shape measurements \citep[from Kilo Degree Survey imaging:][]{DeJong2013,Georgiou2019a}, sample characteristics, covariance estimation etc., can be found. The observed and absolute magnitude distributions of galaxies in GAMA are compared with those from PAUS, for red and blue galaxies, in Fig. \ref{pau:fig:gama_pau_magnitudes}.

We list some summary characteristics for our PAUS W3 and GAMA samples in Table \ref{pau:tab:samples}, including sample counts, mean redshifts and mean luminosities, relative to a pivot luminosity $L_{\rm{piv.}}$ corresponding to absolute $M_{r}=-22$ -- this is for comparison with literature studies of IA \citep[\eg][]{Joachimi2011,Johnston2019,Fortuna2020}.

\section{Random galaxy catalogues}
\label{pau:sec:randoms}

Configuration-space statistics involving galaxy positions typically rely upon sets of random points as Monte Carlo samples of the observed survey volume. Galaxy densities are taken in ratio to the density of these un-clustered points, which grant a notion of the local `mean' density. In addition, the subtraction of statistics measured around random points, from those measured around galaxies, helps to mitigate biases coming from survey edge effects and spatially correlated systematic effects in studies of, \emph{e.g.}, intrinsic alignments and galaxy-galaxy lensing \citep{Singh2016}.

Our objective here is to explore the clustering and IA of PAUS galaxies over short separations in three dimensions, thus our randoms need to reproduce both the radial and angular selection functions of the data, without structure and at high resolution. The latter selection function is trivially reproduced, as we are able to construct an angular mask from observed galaxy positions, within which we can assign uniformly distributed on-sky positions to random points -- thus replicating the survey window and stellar masking from Fig. \ref{pau:fig:W3footprint}.

The radial selection function is more difficult to characterise, as simple fits to, or reshuffling of, the galaxies' $n(z)$ would retain structural information and act to erase line-of-sight correlations. Moreover, we intend to make sample selections in the galaxy data to capture the galaxy type-dependence (and, in future work, any redshift-evolution, luminosity dependence etc.) of IA/clustering phenomena -- these selections will change the galaxies' $n(z)$ and must be reflected in the radial distribution of randoms. To tackle these challenges, we chose to follow and adapt the galaxy-cloning method introduced by \cite{Cole2011} and employed in the GAMA survey clustering analysis of \cite{Farrow2015}.

\subsection{$V_{\rm{max}}$ randoms}
\label{pau:sec:vmax_randoms}

The method uses indirect estimates of the galaxy luminosity function (LF) to predict an un-clustered $n(z)$. Given the survey limiting characteristics and galaxy properties, one can compute for each galaxy the maximum volume \vmax within which it could be observed -- this corresponds directly to a maximum redshift {$z_{\rm{max}}$}, dependent upon the survey flux-limit and galaxies' $k+e$-corrected magnitudes in the relevant detection band. $k$-corrections modify magnitudes to account for redshifting of SEDs, whilst $e$-corrections account for their evolution; to estimate the magnitude of an object were it at $z=0$, one must consider that the object's SED would be more evolved in this case, hence the observed magnitudes need an additional correction. We computed $z_{\rm{max}}$ via the relations

\begin{equation}
    \begin{split}
        M_{z=0} = & \quad m_{\rm{obs.}} - \mu_{\rm{obs.}} - k_{0}(z_{\rm{obs.}}) + Q\,z_{\rm{obs.}} \\
        = & \quad m_{\rm{limit}} - \mu_{\rm{max}} - k_{0}(z_{\rm{max}}) + Q(z_{\rm{obs.}} - z_{\rm{max}}) \quad , \\
        \label{pau:eq:zmax_relation}
    \end{split}
\end{equation}
where $M_{z=0}$ denotes the rest-frame absolute magnitude at $z=0$, $m$ are observed/limiting apparent magnitudes, $\mu$ are the observed/maximum distance moduli\footnote{Not to be confused with the multiplicative ellipticity bias from Sec. \ref{pau:sec:shape_calibration}.}, $k_{0}(z)$ are $k$-corrections from redshift $z$ to zero, and $Q(z - z_{\rm{ref}}) \equiv e(z)$ parameterises the evolution correction between a galaxy redshift $z$ and a reference redshift $z_{\rm{ref}}$ -- assuming galaxy magnitudes to evolve linearly with redshift. This parameterisation is highly approximate, and will not correctly capture the complex evolution of stellar populations over cosmic time; a more comprehensive treatment could make use of spectral synthesis models, though that is beyond the scope of this work.

The maximum volume is

\begin{equation}
    V_{\rm{max}} = \int^{z_{\rm{max}}}_{z_{\rm{min}}} \frac{{\rm{d}}V}{{\rm{d}}z} \, {\rm{d}}z \quad ,
    \label{pau:eq:vmax}
\end{equation}
where we have not yet included the effects of redshift evolution (see \citealt{Cole2011}). $z_{\rm{min}}$ allows for a possible bright cut-off, beyond which highly luminous galaxies are discarded. A standard estimator \citep{Eales1993} for the LF is

\begin{equation}
    \phi(L) = \sum_{i} \, \frac{1}{V_{{{\rm{max}},i}}(L)} \quad ,
    \label{pau:eq:luminosityfunction}
\end{equation}
that is to say, the inverse-$V_{\rm{max}}$ weighted sum over galaxies with luminosity $L$ -- we can thus use our computed \vmax per galaxy to create a randoms catalogue, without radial structure and with a consistent LF, simply by cloning each real galaxy many ($N_{\rm{clone}}$) times and scattering them uniformly within their respective {$V_{\rm{max}}$}.

However, this estimator is vulnerable to bias; galaxy surveys are subject to sampling variance, and thus exhibit significant over/underdensities in the radial dimension -- these will translate into over-/under-representation of luminosity populations. An equal number of clones per galaxy would then over-fit the galaxy $n(z)$, and suppress any measured galaxy clustering. \cite{Cole2011} introduced a maximum likelihood estimator for the LF, computed with a {density-corrected} \vmaxdc which acts to down-weight the contribution of galaxies in overdense environments, \eg clusters. \vmaxdc is computed as

\begin{equation}
    V_{{\rm{max,dc}}} = \int^{z_{\rm{max}}}_{z_{\rm{min}}} \Delta(z) \, \frac{{\rm{d}}V}{{\rm{d}}z} \, {\rm{d}}z \quad ,
    \label{pau:eq:vmaxdc}
\end{equation}
where $\Delta(z)$ is the fractional overdensity as a function of redshift. \vmaxdc can be substituted into Eq. \ref{pau:eq:luminosityfunction} to yield a more robust estimate of the luminosity function. Following \cite{Cole2011} and \cite{Farrow2015}, we use the individual ratios of \vmax to \vmaxdc to re-scale the number of clones per galaxy $n = N_{{\rm{clone}}} V_{{\rm{max}}} / V_{{\rm{max,dc}}}$, such that they are {over}-produced in {under}dense environments, and vice-versa. 

$\Delta(z)$ is estimated through an iterative process, starting with $\Delta(z)\equiv1$ (\ie $V_{{\rm{max}}} \equiv V_{{\rm{max,dc}}}$). We scatter $N_{\rm{clone}}$ clones per galaxy, uniformly throughout their respective {$V_{\rm{max}}$}, and estimate $\Delta(z)$ as

\begin{equation}
    \Delta(z) = N_{{\rm{clone}}} \, \frac{n_{{\rm{gal}}}(z)}{ n_{{\rm{rand.}}}(z)} \quad .
    \label{pau:eq:deltaz}
\end{equation}
One then re-computes {$V_{\rm{max,dc}}$}, re-weights the number of clones, re-generates the randoms' $n(z)$, and repeats the process until $\Delta(z)$ converges -- we allowed for 15 iterations, though convergence was typically reached after $\lesssim10$. Creating randoms for $Q\in[\,-1.5,1.5\,]$, with a spacing of $0.1$, we found on inspection of the PAUS W3 $n(z_{\rm{phot.}})$ that the $Q=0.2$ randoms provide the best match to the redshift distribution -- we leave a more detailed optimisation of $Q$, or a more complex treatment of $e$-corrections, to future work.

\subsection{Windowing}
\label{pau:sec:windowing}

Thus far, we have ignored the redshift evolution of the luminosity function. \cite{Cole2011} outlines methodology to include a parameterised model for the evolution, to be constrained by the data. To avoid the need for some parametric LF-evolution model, we implement the `windowing' alternative employed by \cite{Farrow2015}, wherein each galaxy's \vmax is limited by a window function centred on its observed radial position. Galaxy clones are then scattered only within the window, limiting their presence in disparate redshift regimes and thus building the $z$-evolution into the randoms.

Each window takes a Gaussian form, truncated at $\pm2\sigma$ such that $\sim71.5\%$ of a galaxy's clones will exist within a $1\sigma$ \emph{volume} deviation from its observed redshift -- this means that the window is elongated towards the observer in the radial dimension, as the observed volume diminishes due to the light-cone effect. It is necessary that each window be reflected at any boundaries -- \ie $z=0$, $z_{\rm{min}}$, $z_{\rm{max}}$ or the limits of the survey -- in order to prevent an excessive stacking-up of clones in the centre of the randoms' redshift distribution.

The choice of width $\sigma$ for the Gaussian is fairly arbitrary, however it clearly must be large enough that radial large-scale structure is not over-fitted, and small enough to achieve the desired preservation of luminosity populations over redshifts in the clones. We produce windowed randoms for PAUS with $\sigma=6\times10^{6}\,(h^{-1}{\rm{Mpc}})^{3}$, chosen to satisfy these requirements. The window function, recast as a function of redshift/comoving distance, is included in the integrals of \vmax and \vmaxdc (Eqs. \ref{pau:eq:vmax} \& \ref{pau:eq:vmaxdc}) so that clone-counts are correctly adjusted for the new, Gaussian-weighted volumes.

\begin{figure}
    \centering
    \includegraphics[width=\columnwidth]{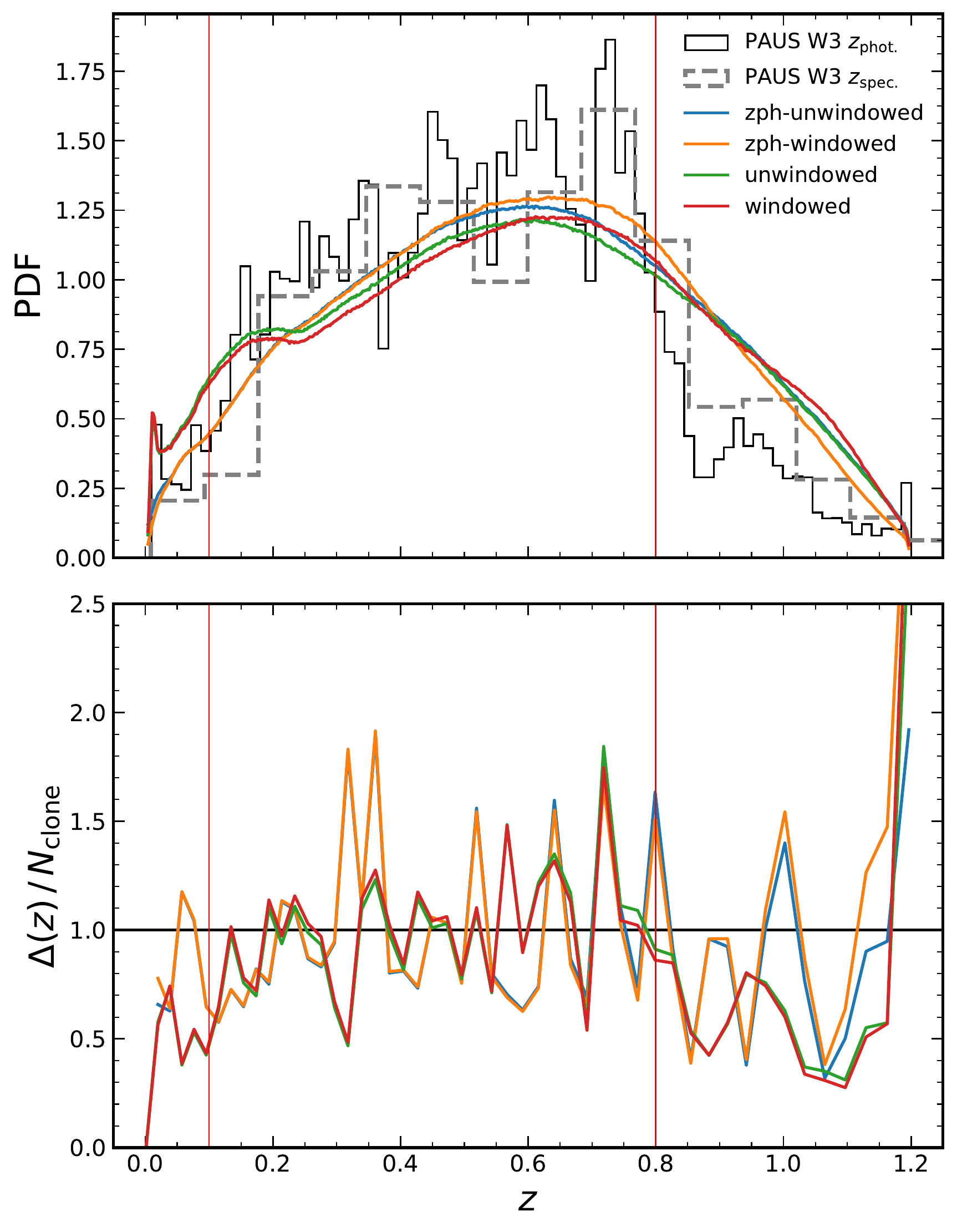}
    \caption[PAUS: the redshift distributions of PAUS W3 galaxies and our un/windowed random galaxy clones]{\emph{Top:} Photometric redshift distribution of PAUS W3 (black solid), with spectroscopic redshifts of the DEEP2-matched galaxies (grey dashed), overlain with the randoms' $n(z)$, generated with and without the windowing (Sec. \ref{pau:sec:windowing}) and \pz (`zph-randoms'; Sec. \ref{pau:sec:photoz_randomsimpact}) approaches, as indicated in the legend. \emph{Bottom:} Density contrast $\Delta(z)$ (Eq. \ref{pau:eq:deltaz}) computed after the final iteration of clone dispersal for each set of randoms -- for zph-randoms, the ratio $n_{\rm{gal}}(z)\,/\,n_{\rm{rand.}}(z)$ (Eq. \ref{pau:eq:deltaz}) is computed against galaxy redshifts drawn from $n(z_{\rm{spec.}}\,|\,z_{\rm{phot.}})$ (see Sec. \ref{pau:sec:photoz_randomsimpact}), such that $n_{\rm{gal}}(z)\neq{}n(z_{\rm{phot.}})$ and $\Delta(z)\,/\,N_{\rm{clone}}$ can be greater than unity where $n_{\rm{rand.}}(z)>n(z_{\rm{phot.}})$, \eg at $z_{\rm{phot.}}\sim1$. Our zph-randoms methods yield smoother redshift distributions, more faithful to the available $n(z_{\rm{spec.}})$. \tbf{Red vertical lines delimit the redshift range employed in our correlation function analysis.}
    }
    \label{pau:fig:randoms}
\end{figure}

\begin{figure}
    \centering
    \includegraphics[width=\columnwidth]{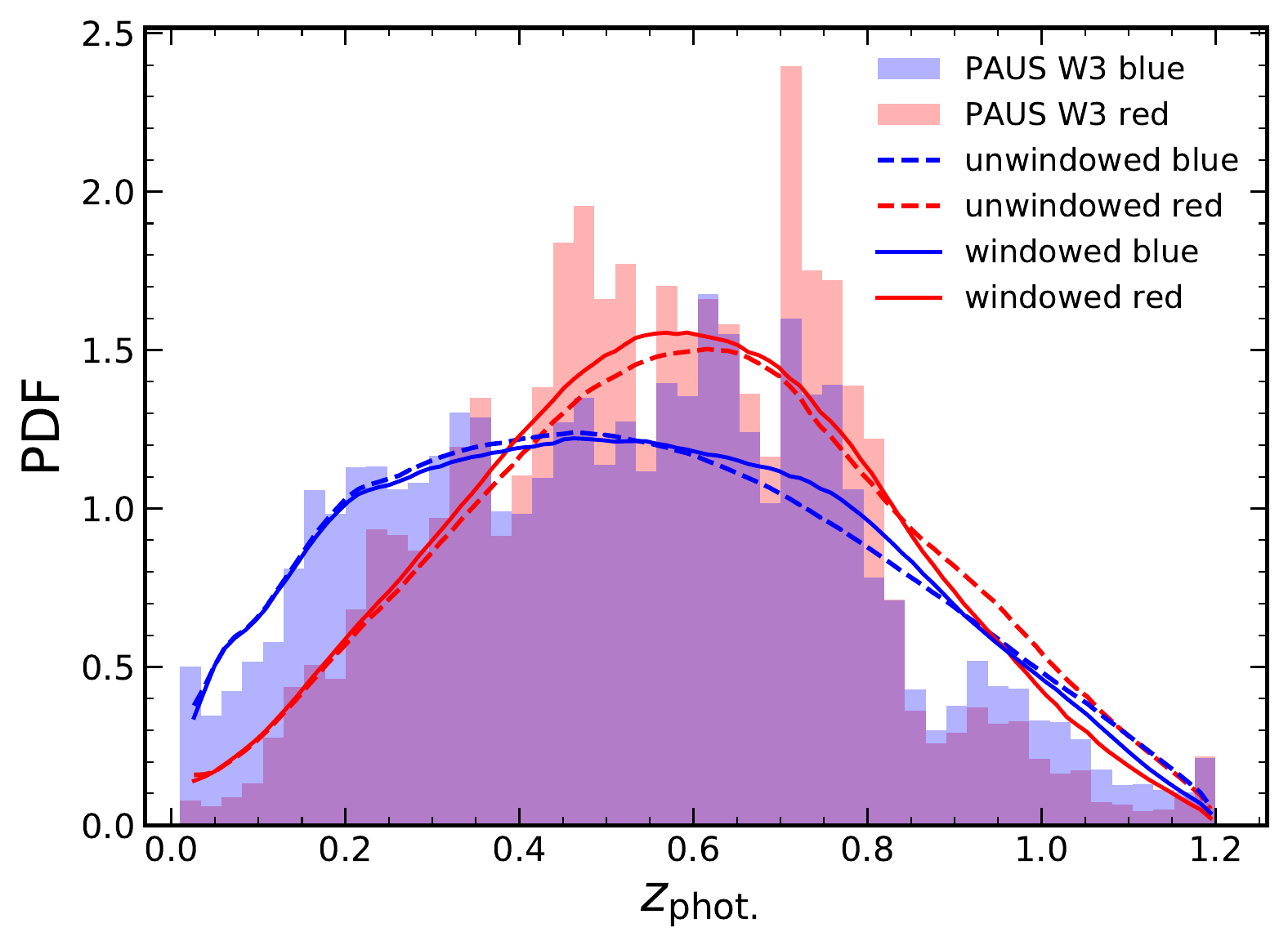}
    \caption[PAUS: redshift distributions of red/blue PAUS galaxies compared with those of the relevant random galaxy clones]{Photometric redshift distributions of PAUS W3 galaxies, split into ({\sc{LePhare}}) red/blue samples (shaded histograms), along with the $n(z)$ for red/blue clones in the windowed randoms (solid lines) and the unwindowed randoms (dashed lines). Randoms shown here are the `zph-randoms' described in Sec. \ref{pau:sec:photoz_randomsimpact}. Each histogram/curve is individually normalised to unit area.
    }
    \label{pau:fig:rb_randoms}
\end{figure}

Figure \ref{pau:fig:randoms} shows the PAUS 40-NB photometric redshift distribution, with the spectroscopic redshifts of available DEEP2 galaxies, and overlaid with redshift distributions for our un/windowed, cloned randoms. The bottom panel gives the overdensity $\Delta(z)$ as a function of redshift. To illustrate the utility of galaxy cloning, Fig. \ref{pau:fig:rb_randoms} also shows the photo-$z$ distributions of ({\sc{LePhare}}) red and blue galaxies in PAUS W3 (see Fig. \ref{pau:fig:PAUGAMAzCMD}), along with those of their relevant random clones. One sees that the general trends levied by the selection are reproduced by both un/windowed randoms, though the windowing restriction causes differences, especially for the red galaxy randoms.

\begin{figure}
    \centering
    \includegraphics[width=\columnwidth]{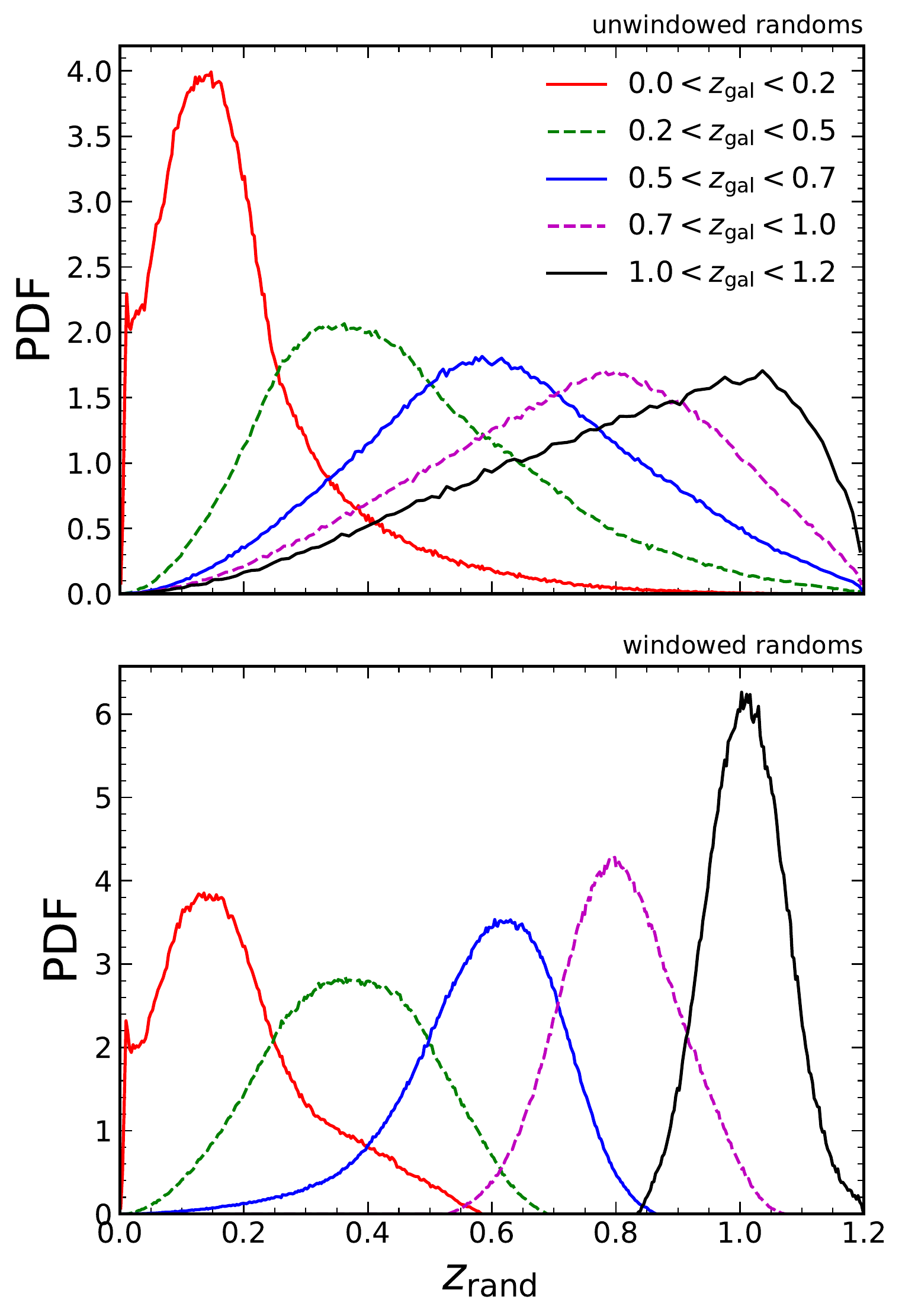}
    \caption[PAUS: redshift distributions of un/windowed random clones corresponding to PAUS galaxies binned in redshift]{Redshift distributions of random clones $z_{\rm{rand}}$, whose parent galaxies are situated at photometric redshifts $z_{\rm{gal}}$. \emph{Top:} Unwindowed clones are scattered over the entire redshift range, depending on the brightness, and hence $z_{\rm{max}}$, of parent galaxies. \emph{Bottom}: For the windowed randoms, 71.5\% of a galaxy's random clones are scattered within a $\pm{1}\sigma$ symmetric volume, centred on the location of the parent galaxy [we display $\sigma=3\times10^{6}\,(h^{-1}{\rm{Mpc}})^{3}$ randoms here, for a clearer illustration]. We note that the symmetry is in volume coordinates, hence in redshift/comoving coordinates the windows are extended in the direction of the observer, and slightly squashed in the outward direction. This figure is closely based on a similar plot presented by \cite{Farrow2015} for GAMA galaxies and randoms (their Fig. 3). Plateaus in the red curves, on approach to redshift zero, correspond to faint, very low-redshift galaxies which are excluded from our analysis.
    }
    \label{pau:fig:zbin_windows}
\end{figure}

\begin{figure}
    \centering
    \includegraphics[width=\columnwidth]{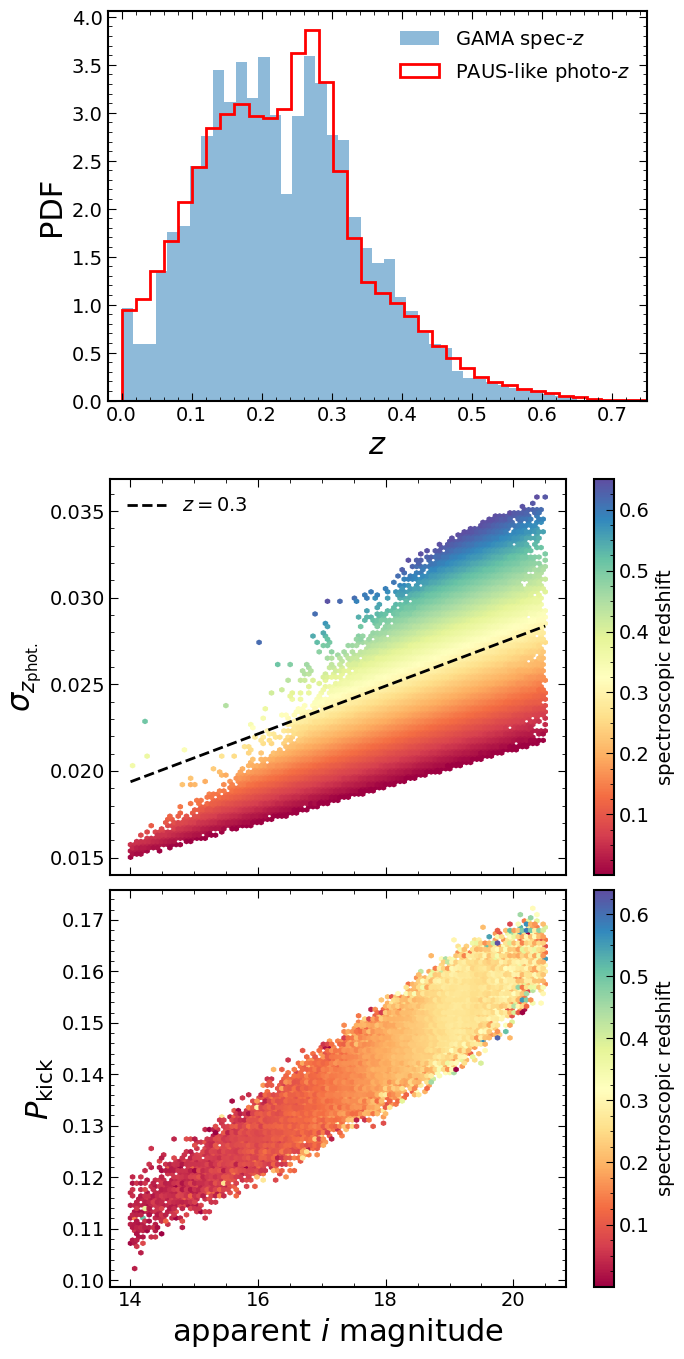}
    \caption{GAMA spectroscopic redshift distribution (\emph{top-panel;} blue), overlaid with our approximately PAUS-like photometric redshift distribution (\emph{top-panel;} red) -- these photo-$z$ are created by applying a redshift- and magnitude-dependent Gaussian scatter (\emph{middle-panel}; Eq. \ref{pau:eq:pauslike_photoz_scatter}) to the spec-$z$, along with a probabilistic `kick' (\emph{bottom-panel}; Eq. \ref{pau:eq:pauslike_photoz_Pkick}) to generate catastrophic outliers, and a manual relocation of some galaxies to $z_{\rm{phot.}}\sim0.27$ (see Sec. \ref{pau:sec:photoz_randomsimpact}). Bottom-panels are hexagonally-binned 2D histograms, with cells coloured according to the mean GAMA spec-$z$ of objects residing in each cell. %The $\sigma_{z_{\rm{phot.}}}$ vs. $i$ magnitude relation is shown for a fixed redshift $z=0.3$, as a black dashed line.
    }
    \label{pau:fig:pauslikephotozingama}
\end{figure}

The effects of windowing are illustrated in Fig. \ref{pau:fig:zbin_windows}, which closely mimics Fig. 3 from \cite{Farrow2015}. The top-panel shows the unwindowed random clones' redshift distributions for \pz selections in PAUS W3, while the bottom-panel shows the same for $\sigma=3\times10^{6}\,[h^{-1}{\rm{Mpc}}]^{3}$ windowed clones -- we show a smaller window-size in Fig. \ref{pau:fig:zbin_windows} for emphasis of the windowing effect. One clearly sees the restriction of clones to redshifts near their parent galaxy. The numerous clones at very low redshifts (red curves plateau towards $z=0$) correspond to faint, near-universe galaxies with small values of ${z_{\rm{max}}}$ -- these are excluded from our analysis in any case.

\subsection{Photo-$z$ impact \& `zph-randoms'}
\label{pau:sec:photoz_randomsimpact}

As previously mentioned, Fig. \ref{pau:fig:zphot_zspec_qz} reveals a dip in photo-$z$ accuracy whereby many PAUS-DEEP2 galaxies at $z_{\rm{spec.}}\gtrsim0.7$ are assigned $z_{\rm{phot.}}\sim0.7$ by the {\sc{bcnz2}} algorithm -- this is currently under investigation. In the meantime, we note the consequences in Figs. \ref{pau:fig:randoms} \& \ref{pau:fig:rb_randoms}, where each set of randoms seemingly overpopulates redshifts $z\gtrsim0.8$, relative to the PAUS W3 $n(z_{\rm{phot.}})$. This is because (i) the number of PAUS galaxies beyond $z_{\rm{phot.}}\sim0.8$ should be higher (Fig. \ref{pau:fig:zphot_zspec_qz}), and (ii) the inferred \zmax are biased for many galaxies with $z_{\rm{phot.}}\sim0.7$, due to significant errors in $z_{\rm{phot.}}$ (\ie in the observed redshift $z_{\rm{obs.}}$ in Eq. \ref{pau:eq:zmax_relation} -- these biases are also responsible for the low-redshift `bumps' exhibited by randoms in Fig. \ref{pau:fig:randoms}, where galaxies' \zmax have been underestimated). Consequently, the randoms `see' a large underdensity at $z>0.8$ and respond by over-filling the volume with clones. The windowed randoms restrict the clones to redshifts near to their parent $z_{\rm{phot.}}$, resulting in the marginally tighter fit to the data in Fig. \ref{pau:fig:rb_randoms}.

Using GAMA as a test-bed, we investigated the photo-$z$ degradation of measurable signals, and the consequences of generating randoms using photo-$z$. In order to roughly mimic the pathologies present in the PAUS photo-$z$ distribution, we applied a Gaussian\footnote{Modelling \pz errors as Gaussian-distributed is perhaps not the most accurate way to characterise typical \pz distributions with significant proportions of catastrophic outliers; we do so here as our mock GAMA samples are not intended to be completely realistic but rather instructive. A more detailed application could explore, for example, the student-t distribution, the thicker tails of which were found to be a better descriptor for KiDS luminous red galaxy \pz scatter by \cite{Vakili2020}.} scatter $\sigma_{z_{\rm{phot.}}}$ to GAMA spec-$z$ according to

\begin{equation}
    \sigma_{z_{\rm{phot.}}} = 0.02 \, \frac{i}{i_{50}} \, (1+z_{\rm{spec.}}) \quad ,
    \label{pau:eq:pauslike_photoz_scatter}
\end{equation}
where $i$ is the observed $i$-band magnitude of a galaxy, and $i_{50}$ is the $50^{\rm{th}}$ percentile of all $i$-magnitudes in the range $14<i<20.5$ -- thus fainter objects suffer larger scatters. We further applied a probabilistic `kick' to GAMA spec-$z$ in order to mimic photo-$z$ outliers. The probability $P_{\rm{kick}}$ of a galaxy receiving a kick is implemented as

\begin{equation}
    P_{\rm{kick}} = 0.15 \, \frac{i}{i_{50}} + \mathcal{N}\left[0, 0.003\right] \quad ,
        \label{pau:eq:pauslike_photoz_Pkick}
\end{equation}
where the Gaussian draw introduces stochasticity about the relation, and the kick $\delta{z}$ itself is uniformly drawn from $0.07<\delta{z}<0.08$ before a random 60\% are given a negative sign -- such that model catastrophic photo-$z$ failures tend to {under}estimate the true redshifts (Fig. \ref{pau:fig:zphot_colour}). $\sigma_{z_{\rm{phot.}}}$ and $P_{\rm{kick}}$ are illustrated in the bottom-panels of Fig. \ref{pau:fig:pauslikephotozingama}, as functions of the $i$-band magnitude, and coloured by the GAMA spectroscopic redshifts of objects. Finally, to model the stacking-up of photo-$z$ estimates at $z_{\rm{phot.}}\sim0.7$ in PAUS, we re-located to $z_{\rm{phot.}}=0.27+\mathcal{N}\left[0,0.025\right]$ (a) a random 3\% of all GAMA spec-$z$, and (b) a number of galaxies around $z_{\rm{spec.}}\sim0.35$, with the probability of relocation equal to 40\% at the $z_{\rm{spec.}}=0.35$ peak, and decaying as a Gaussian ($\sigma=0.025$) to either side.

The resulting redshift distribution (Fig. \ref{pau:fig:pauslikephotozingama}, top-panel) is thus smoothed, as is typical for photo-$z$ distributions, and features a peak at $z_{\rm{phot.}}\sim0.27$ followed by a sharp drop -- these redshifts provide our PAUS-like\footnote{As Fig. \ref{pau:fig:zphot_colour} shows, photo-$z$ in PAUS are variably precise for red and blue galaxies -- we do not attempt to reproduce these trends in our mock GAMA photo-$z$, leaving a more robust mimicry of PAUS photo-$z$ -- \eg featuring \pz errors drawn from a joint probability distribution describing the \pz bias given spec-$z$, colours, magnitudes, sizes etc. -- to future work.} photo-$z$ testing ground for a new method of redshift-assignment for random points: \tbf{employing galaxy redshifts sampled from conditional distributions $n(z_{\rm{spec.}}\,|\,z_{\rm{phot.}})$, rather than \pz estimates themselves}. The aforementioned biases in $z_{\rm{max}}$ estimates, due to the inaccuracies inherent to photo-$z$, can be mitigated over the galaxy ensemble if we draw many realisations of a `spectroscopic' $n(z)$ from the \tbf{probability density} distributions $n(z_{\rm{spec.}}\,|\,z_{\rm{phot.}})$ surrounding each galaxy's $z_{\rm{phot.}}$ -- after testing, we settled upon 320 draws per object, from $n(z_{\rm{spec.}}\,|\,z_{\rm{phot.}}\pm0.03)$ for GAMA, and we increased this to 500 draws from the conditional range $z_{\rm{phot.}}\pm0.04$ for PAUS. The resulting distributions of $z_{\rm{max}}$ are illustrated in Appendix Fig. \ref{pau:fig:zspeczmax_draws} for a random selection of GAMA galaxies.

\begin{figure}
    \centering
    \includegraphics[width=\columnwidth]{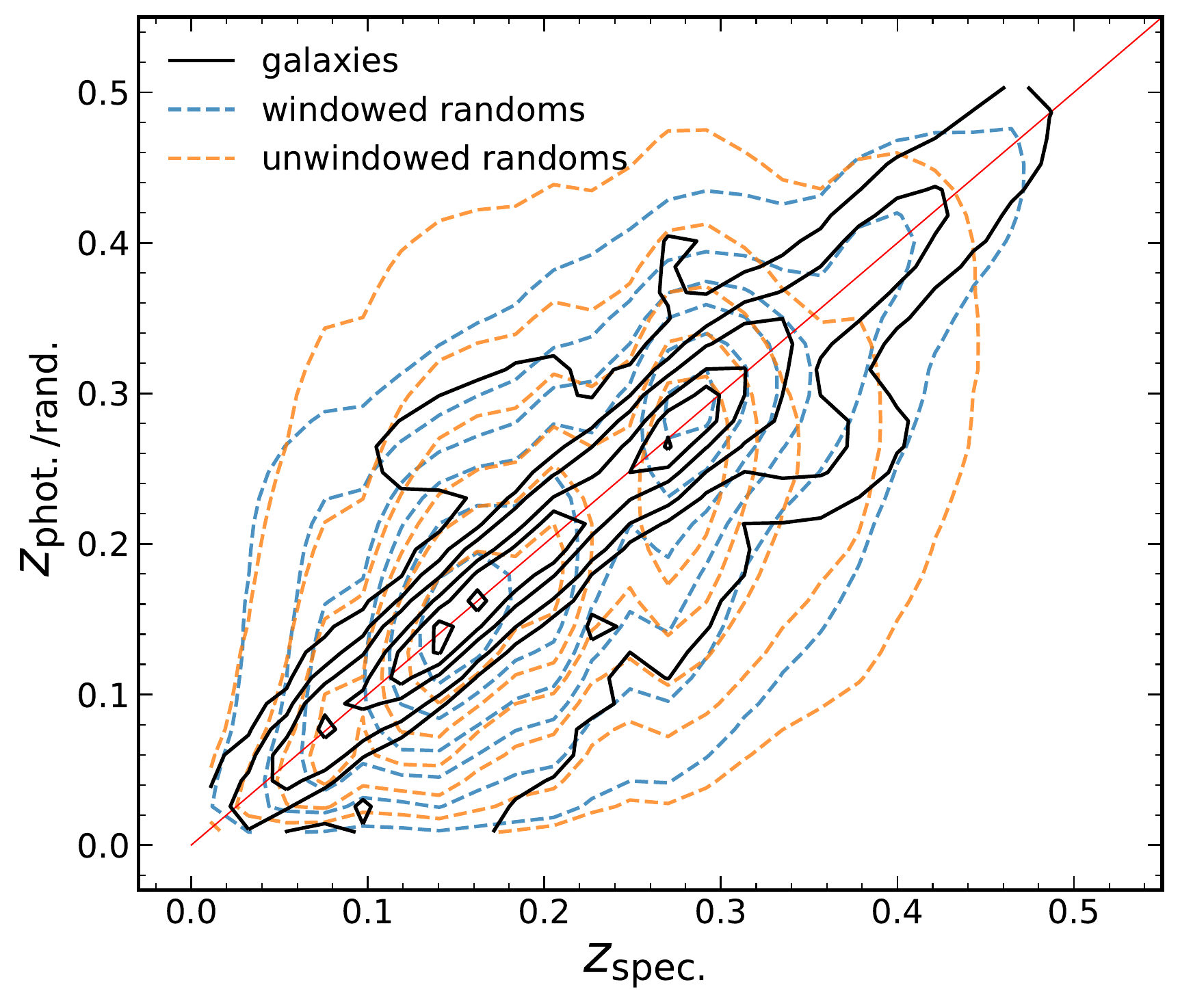}
    \caption{Contours depicting cumulative population fractions $0.08,\,0.25,\,0.42,\,0.58,\,0.75$ and $0.92$ for GAMA galaxy/zph-randoms catalogues on the $z_{\rm{phot./rand.}}$ vs. $z_{\rm{spec.}}$ plane, where the y-axis refers to $z_{\rm{phot.}}$ for galaxies and $z_{\rm{rand.}}$ for randoms. For randoms, $z_{\rm{spec.}}$ refers to the spectroscopic redshift of the parent galaxy to a clone at redshift $z_{\rm{rand.}}$. Blue contours depict our windowed randoms, with window-width $\sigma=5\times10^6\,(h^{-1}\rm{Mpc})^3$, and orange contours are for unwindowed randoms. One clearly sees the intentional photo-$z$ biases (Sec. \ref{pau:sec:photoz_randomsimpact}) in the galaxy contours (solid black), and how the randoms (dashed colours) softly mimic them. Fig. \ref{pau:fig:nzszp_marginals} shows five `marginals' through this distribution for (zph-windowed randoms only), collapsing the y-axis to show $n(z_{\rm{spec.}}\,|\,z_{\rm{phot.}})$.
    }
    \label{pau:fig:nzszp_contours}
\end{figure}

\begin{figure*}
    \centering
    \includegraphics[width=0.8\linewidth]{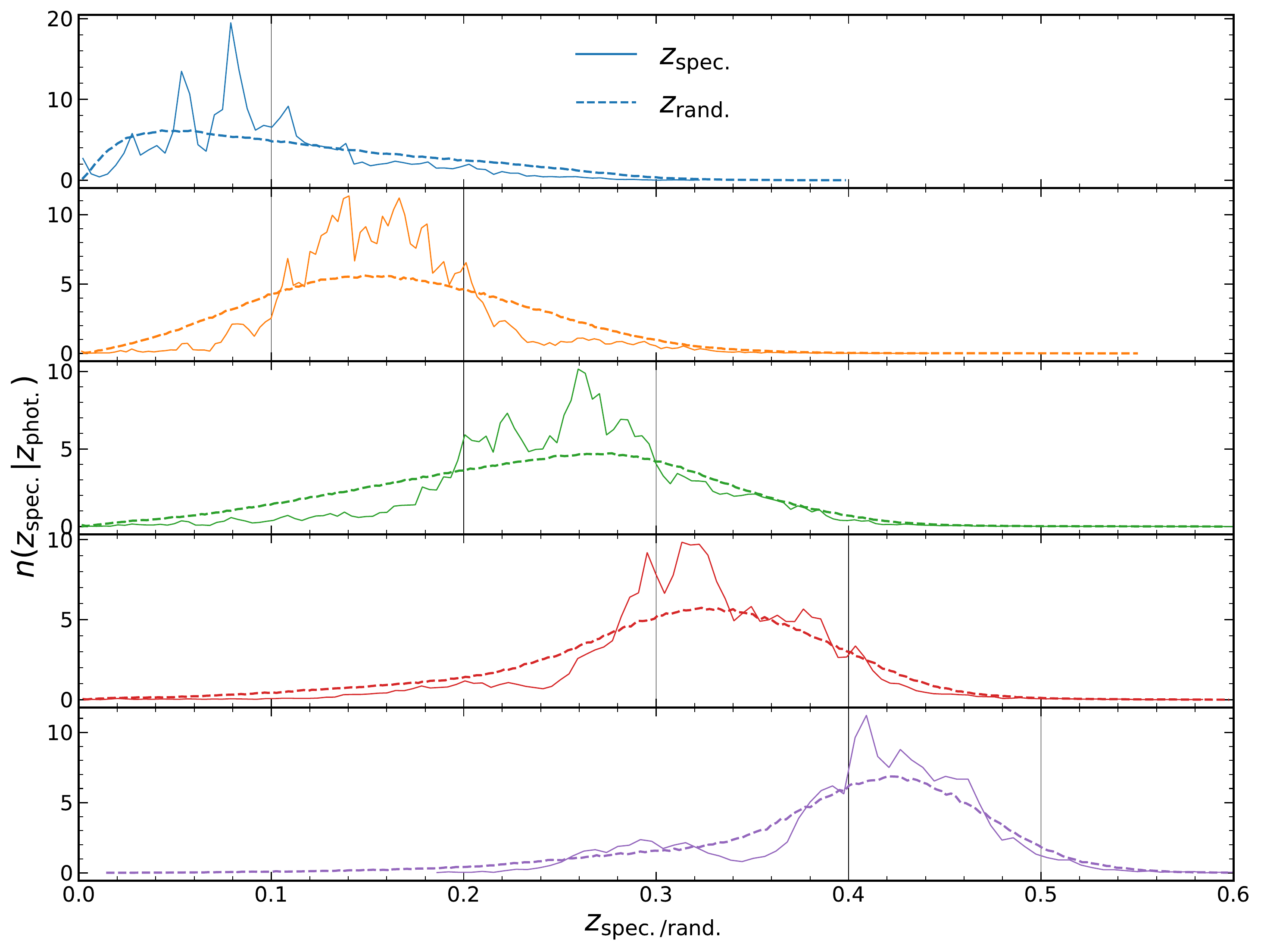}
    \caption{Spectroscopic redshift distributions of GAMA galaxies, binned by the mock photometric redshifts we describe in Sec. \ref{pau:sec:photoz_randomsimpact}. Vertical black lines depict the bin-edges in $z_{\rm{phot.}}$ and solid curves give the resulting $n(z_{\rm{spec.}}\,|\,z_{\rm{phot.}})$ for galaxies. Selecting only the clones of those binned galaxies, from our `zph-windowed' GAMA randoms [$\sigma=5\times10^{6}\,(h^{-1}{\rm{Mpc}})^{3}$], the randoms' redshift distributions are then given by the dashed curves. One sees that our `zph-randoms' [generated using samples from the $n(z_{\rm{spec.}}\,|\,z_{\rm{phot.}})$'s centred on individual galaxies' $z_{\rm{phot.}}$ -- see Sec. \ref{pau:sec:photoz_randomsimpact}] are thus able to trace the artificial photo-$z$ biases -- for example, the secondary peak at $z\sim0.3$ in the final panel, which is captured by the tail of the randoms' $n(z)$. These curves are equivalent to horizontal bands in Fig. \ref{pau:fig:nzszp_contours}, summed over the y-axis.
    }
    \label{pau:fig:nzszp_marginals}
\end{figure*}

Repeating our randoms-generation procedure from Secs. \ref{pau:sec:vmax_randoms} \& \ref{pau:sec:windowing} for each of the 320 realisations of GAMA, we created ensemble-sets of randoms which encode the characteristic errors in the photo-$z$ distribution, as traced by the available \protect{spec-$z$\footnote{The conditional $n(z_{\rm{spec.}}\,|\,z_{\rm{phot.}})$ is formed only by galaxies for which we have both spec-/photo-$z$ estimates; that is to say, all galaxies in GAMA, but only DEEP2 galaxies in PAUS. Thus the equivalent procedure in PAUS will be more exposed to biases stemming from sample variance. An interesting route to minimise such biases would be to use individual galaxy redshift probability density estimates $P(z)$ to construct $n(z_{\rm{spec.}}\,|\,z_{\rm{phot.}})$. This would require a degree of testing -- \eg running {\textsc{bcnz2}} against GAMA to test the fidelity of $P(z)$'s -- that we leave to future work.}}. We illustrate the outcomes of this procedure in Figs. \ref{pau:fig:nzszp_contours} \& \ref{pau:fig:nzszp_marginals}, which display the $n(z_{\rm{spec.}})$ vs. $n(z_{\rm{phot.}})$ plane for galaxies/randoms, and some `marginals' through the plane, respectively. One sees from the contour-plot (Fig. \ref{pau:fig:nzszp_contours}) that the deviations from 1-to-1 correspondence in the data are smoothly traced by the randoms, and this is made clearer in Fig. \ref{pau:fig:nzszp_marginals}, where the hand-made peak at $z_{\rm{phot.}}\sim0.27$ (solid line) is captured by our ensemble [$\sigma=5\times10^{6}\,(h^{-1}{\rm{Mpc}})^{3}$] windowed randoms (dashed line) in the final panel.

To assess the performance of these `zph-randoms', we compare intrinsic alignment and clustering correlations in GAMA, measured using spec-/photo-$z$ and various random galaxy catalogues, presenting results in Appendix \ref{pau:sec:app_zphrandoms}. In particular, we see that this `excess' of randoms at $z\sim0.27$, matching the photo-$z$-induced galaxy excess there, is important; correlated galaxy pairs at these redshifts have some real range of transverse separations $r_{p}$, but are included in 3D correlation function estimators at separations $>r_{p}$ because they are thought to be at higher redshifts due to \pz errors. This causes a tilting of observed correlations (see Fig. \ref{pau:fig:signalbreakdown_zphrandoms}), as small-scale power is erroneously pushed to larger scales -- the excess in the randoms, however, mitigates this effect by suppressing long-range correlations. Next, we detail our methodology for measuring projected correlations.

\section{Two-point statistics}
\label{pau:sec:method}

With random galaxies uniformly permeating the survey volume, we are now free to measure the galaxy clustering and alignments with a radial binning of galaxy pairs. Since PAUS is a unique survey, we lack samples against which to directly compare galaxy statistics.

Photometric redshift scatter acts as a radial smoothing of the 3D galaxy density field, suppressing the observable clustering as galaxy pairs which are not in fact correlated pollute the desired signal. Thus we expect any clustering signal from PAUS to be lower in amplitude than the equivalent signal from a spectroscopically observed sample. One expects a similar dilution for the intrinsic alignment signal, as uncorrelated galaxy pairs are mistakenly included in the estimator.

As such, we choose to compare our PAUS signals with those measured in GAMA using our PAUS-like mock photo-$z$. We stress that this procedure is highly approximate, and intended only to be instructive -- \pz scatter is often a function of galaxy properties, which are in turn correlated with environments. Thus our degradation of GAMA redshifts, whilst reminiscent of PAUS over the full $n(z)$, may not match the severity or complexity of the degradation already present in PAUS. PAUS is also deeper than GAMA, and has less area; the signal-to-noise will peak in a different regime. For these reasons, the signal-comparison ought not to be considered rigorous.

\subsection{Clustering}
\label{pau:sec:clust_method}

The Landy-Szalay \citep{Landy1993} estimator for the galaxy correlation function is

\begin{equation}
    \hat{\xi}_{\rm{gg}}(r_{p}, \Pi) = \frac{D_{i}D_{j} - D_{i}R_{j} - D_{j}R_{i} + R_{i}R_{j}}{R_{i}R_{j}} \Bigg{|}_{r_{p},\Pi} \quad ,
\end{equation}
where the various galaxy-galaxy ($DD$), random-random ($RR$) and galaxy-random ($DR$) pair-counts are binned by transverse \rp and radial $\Pi$ comoving separations. Subscripts $i,j$ label the galaxy ($D$) samples and their corresponding randoms ($R$); $i=j$ for a sample auto-correlation. In this work, we display only the galaxy clustering auto-correlations; blue-blue and red-red sample clustering. This is done to provide a direct comparison between the clustering of red and blue galaxies, though future analyses will make use of full-sample clustering correlations in order to constrain the galaxy bias in tandem with IA correlations. Here, we have used the full (unselected) PAUS W3 sample as a positional tracer for IA correlations, thus improving the signal-to-noise and eliminating the effects of differential galaxy bias in the `density' samples (see Eq. \ref{pau:eq:xigp_estimator_mandelbaum06}).

The projected clustering correlation function is then

\begin{equation}
    \hat{w}_{\rm{gg}}(r_{p}) = \int^{\Pi_{\rm{max}}}_{-\Pi_{\rm{max}}} \hat{\xi}_{\rm{gg}}(r_{p}, \Pi) \, \rm{d}\Pi \quad .
    \label{pau:eq:wgg_xiggintegrated}
\end{equation}
Working with a log-spaced binning in \rp of 5 bins between $0.1 - 18\mpc$, we explored different choices of binning in $\Pi$ to accommodate the effects of photo-$z$ scatter. The standard approach, utilised for spectroscopic data, is to perform a Riemann sum over uniform bins in $\Pi$, \eg with $\rm{d}\Pi=4\mpc$ and $|\Pi_{\rm{max}}|=60\mpc$ \citep[as in \eg][]{Mandelbaum2006,Mandelbaum2011,Johnston2019}. With precise redshift information, these limits capture the vast majority of correlated galaxy pairs without introducing excessive noise due to the inclusion of uncorrelated objects. In the photometric case, however, the signal is smeared-out along the $\Pi$-axis, and correlated pairs are lost from the estimator. We explored a `dynamic' binning in $\Pi$, where bins increase in size from small to large values of $\Pi$ -- thus physically associated objects are brought back into the estimator, and the impacts of integrating over noise at large-$\Pi$ are mitigated with broader bins. We implemented this binning as an adapted Fibonacci sequence\footnote{This choice is arbitrary -- we only require a sequence with a shallower-than-exponential gradient, and the Fibonacci numbers are thus convenient.}, up to a $|\Pi_{\rm{max}}|$ of $233\mpc$, such that the bin-edges are

\begin{equation}
    |\Pi_{\rm{dynamic}}| = 0,1,2,3,5,8,13,21,34,55,89,144,233\mpc \quad .
    \label{pau:eq:fibonacci_bins}
\end{equation}
We shall display signals measured with both the standard, `uniform' and new, dynamic binning in $\Pi$. With this $\Pi$-binning, and our `zph-randoms' (Sec. \ref{pau:sec:photoz_randomsimpact}), we hope to recover as much lost signal as possible, whilst simultaneously mitigating the impacts of photo-$z$ outliers upon the signals, thus simplifying the modelling of correlations in future work.

\subsection{Alignments}
\label{pau:sec:alignments_method}

With the same $r_p,\Pi$ binning choices, we also measured the projected galaxy position-intrinsic shear correlations $w_{\rm{g+}}$ for our PAUS galaxy samples, using the estimator \citep{Mandelbaum2006}

\begin{equation}
    \hat{\xi}_{\rm{g+}}(r_{p},\Pi) = \frac{S_{+}D - S_{+}R_{D}}{R_{S}R_{D}} \, \Bigg{|}_{r_{p},\Pi} \quad ,
    \label{pau:eq:xigp_estimator_mandelbaum06}
\end{equation}
for which
\begin{equation}
    S_{+}D = \sum_{i\neq{}j\,|\,r_{p},\Pi} \frac{\epsilon_{+}(j|i)}{\mathcal{R}} \quad .
    \label{pau:eq:ellipticitycomponents}
\end{equation}
$D$ here denotes the `density' sample of galaxies, which forms the `position' component of the correlation (as mentioned above, this is the full, all-colour galaxy sample for PAUS), $S_{+}$ denotes the `shapes' galaxy sample and $S_{+}X$ is the sum of ellipticity components of galaxies $i$ from the shapes sample, relative to the vectors connecting them to galaxies $j$ from sample $X$, normalised by the \emph{shear responsivity}\footnote{This quantity differs by a factor of 2, according to the definition of the ellipticity; here we derive shear estimates from the polarisation in Eq. \ref{pau:eq:shear_estimate}, 
hence we do not apply a factor 2 to the responsivity -- see \cite{Mandelbaum2006,Reyes2012}.} $\mathcal{R}\approx(1-\sigma^{2}_{\epsilon})$, where $\sigma_{\epsilon}$ is the $S$ sample shape dispersion. The shape- and density-sample randoms are denoted $R_{S}$ and $R_{D}$, respectively, and $R_{S}R_{D}$ are the normalised pair-counts between the different randoms. We note that the standard convention for $w_{\rm{g}+}$ is that positive signals indicate \emph{radially} aligned galaxies\footnote{As opposed to tangential alignments, which are typically signified by positive signals when studying galaxy-galaxy lensing (GGL).}.

Rotating ellipticities by 45 degrees, $\epsilon_{+}\rightarrow\epsilon_{\times}$ in Eq. \ref{pau:eq:ellipticitycomponents}, and we measure the position-shape cross-component $\hat{\xi}_{\rm{g}\times}$ -- a non-vanishing cross-component would indicate some preferred direction of curl in the galaxy shape distribution, breaking parity and thus signalling the presence of systematics in shape estimation. $\hat{w}_{\rm{g+}}$ and $\hat{w}_{\rm{g}\times}$ are constructed from $\hat{\xi}_{\rm{g+}}$ and $\hat{\xi}_{\rm{g}\times}$ in exact analogy to Eq. \ref{pau:eq:wgg_xiggintegrated}. We present the significance of measured $w_{\rm{g\times}}$ correlations in Appendix \ref{pau:sec:app_zphrandoms}.

We note that increasing the maximum permissible line-of-sight separations $\Pi_{\rm{max}}$ (Eq. \ref{pau:eq:fibonacci_bins}) for galaxy pairs risks the contamination of our intrinsic alignment statistics $w_{\rm{g+}},w_{\rm{g\times}}$ by genuine galaxy-galaxy lensing (GGL) signals -- that is, correlations between foreground galaxy positions and background gravitational shears. The amplitude of any such contribution depends upon the width of the true (\ie not photometric) distribution of $\Pi$ across all pairs considered. With $|\Pi_{\rm{max}}|=233\mpc$, any GGL contribution should be strongest from lenses at $\sim233\mpc$ paired with sources at about twice the lens distance, and dropping off at higher redshifts where efficiently lensed sources are excluded by $\Pi_{\rm{max}}$. Photo-$z$ errors could spuriously promote more distant sources into the estimator (Eq. \ref{pau:eq:xigp_estimator_mandelbaum06}), and any coherent tangential shears around the foreground density field would then be negative in the $\xi_{\rm{g+}}$ correlation -- given the quality of PAUS \pz, and given that we limit our analysis to $z_{\rm{phot.}}>0.1$ ($\gtrsim420\mpc$), we expect any GGL contribution here to be weak; future work with more area and greater signal-to-noise can test for contamination of IA correlations by enforcing large line-of-sight separations in Eq. \ref{pau:eq:xigp_estimator_mandelbaum06} (\ie setting some $|\Pi_{\rm{min}}|$), and should consider this possibility when attempting intrinsic alignment model-fitting.

\subsection{Covariance estimation}
\label{pau:sec:covariances}

We estimated errors for the correlations via a delete-one jackknife resampling of the observed volume. Splitting the PAUS W3 footprint into eight roughly equal-area patches, we then defined redshift slices with comparable galaxy counts to create pseudo-independent jackknife sub-volumes. We defined three equally-populated slices of depth $\geq466\mpc$, in order to accommodate the largest allowed line-of-sight separations in the case of our dynamic binning in $\Pi$, resulting in a total of 24 sub-volumes. The covariance is constructed for each correlation function $w$ as

\begin{equation}
    \hat{C}_{\rm{jack.}} = \frac{23}{24} \, \sum_{\alpha=1}^{24} \, \, (w_{\alpha} - \bar{w})(w_{\alpha} - \bar{w})^{\rm{T}} \quad ,
\end{equation}
where $w_{\alpha}$ is the signal measured upon removal of jackknife sub-volume $\alpha$, and $\bar{w}$ is the average of all jackknife signals $w_{\alpha}$. This definition of jackknife sub-volumes with variable redshifts implicitly assumes that any redshift evolution of the signal is small. Given the small area of PAUS W3, we are otherwise unable to define enough jackknife samples to minimise noise in the covariance, without the errors becoming severely inaccurate on the larger scales of interest -- just 24 sub-volumes for 5 data-points is sub-optimal, though errors should then be reliable up to $\sim19\mpc$ for redshifts $0.1<z<0.8$, thus encapsulating our range of considered scales for PAUS. Since the noise will be prominent in the off-diagonals of the covariance, and since we do not intend to perform any detailed line-fitting here, we proceed with the 3D jackknife, noting that a significant redshift-evolution of signals would result in our {over}-estimating the full-sample variance. We do make use of the full covariances in estimating the significance of detection for IA signals, and in fitting simple power-laws to clustering signals, to aid with comparison of different (with respect to colours, $\Pi$-binning, randoms and \pz quality-cuts) correlation configurations -- see Appendix \ref{pau:sec:app_zphrandoms}.

Any spectroscopic GAMA measurements shown are those presented in \cite{Johnston2019}, with jackknife errors estimated from 45 sub-volumes of depth $\geq150\mpc$, and with colour-selections applied to both density and shapes samples for IA correlations. We estimated errors for our mock \pz GAMA samples with 36 sub-volumes of depth $\geq260\mpc$, noting that the dynamic $\Pi$-binning scenario will consequently suffer slightly under-estimated errors, as pairs falling in the final $\Pi$-bin (Eq. \ref{pau:eq:fibonacci_bins}) are imperfectly captured. These signals are meant only to be instructive, hence we proceed as such.

\section{Results}
\label{pau:sec:pauresults}

\begin{figure*}[!htbp]
    \centering
    \includegraphics[width=\linewidth]{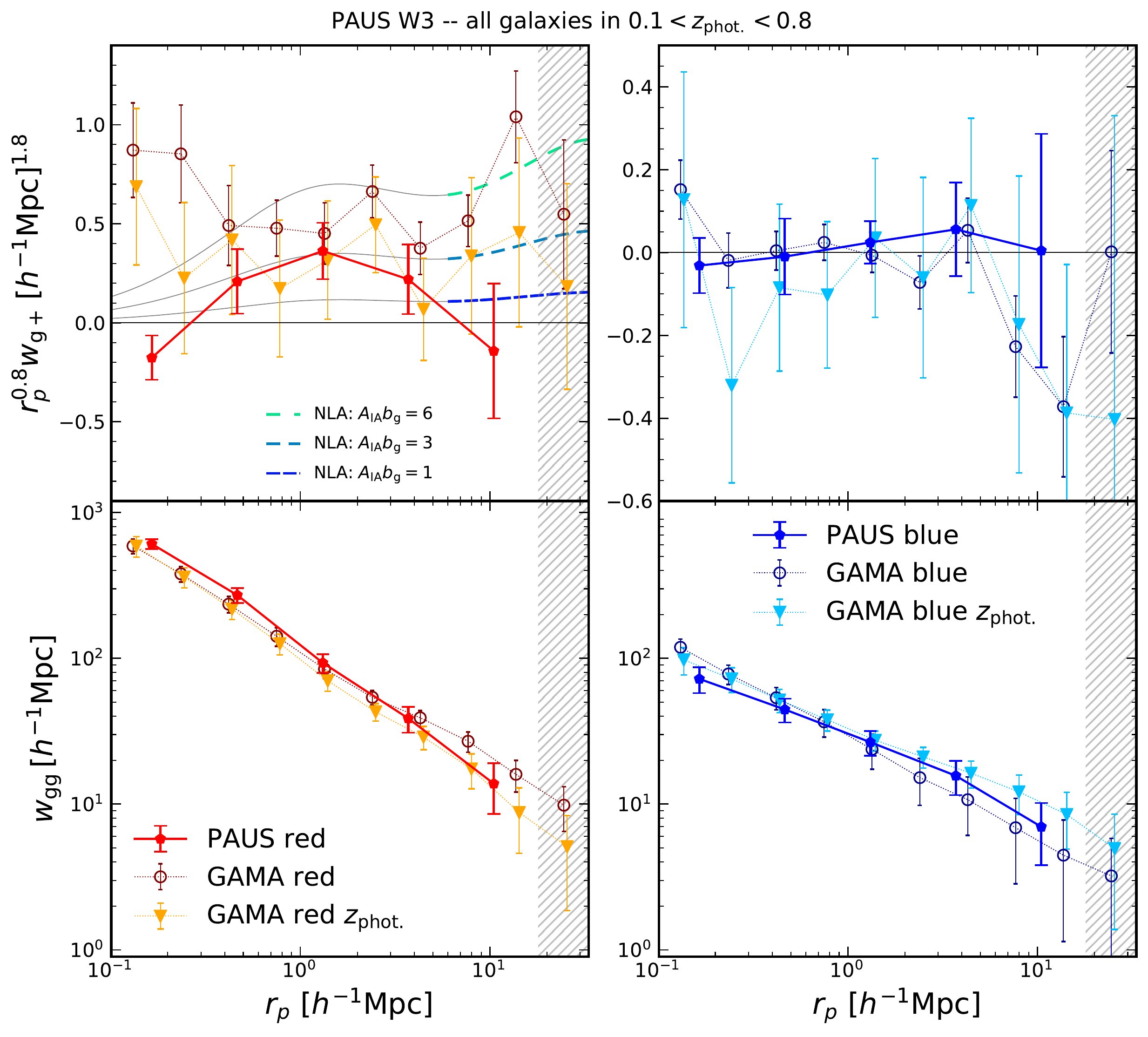}
    \caption[PAUS and GAMA: projected position-intrinsic shear and position-position correlations for red/blue galaxies in PAUS and GAMA]{Projected position-shear (\emph{top}) and position-position (\emph{bottom}) correlations measured for red (\emph{left}) and blue (\emph{right}) galaxies in PAUS (filled pentagons) and GAMA (open circles and filled triangles). The PAUS correlations displayed here are for all galaxies (\ie not selected according to \ttt{Qz}), split by colour according to {\sc{LePhare}}, measured with zph-windowed randoms and dynamic $\Pi$-bins -- see Appendix \ref{pau:sec:app_zphrandoms} for details on our choice here. Downward triangles display the correlations measured in GAMA using our mock photo-$z$, described in Sec. \ref{pau:sec:photoz_randomsimpact}. Grey hatching indicates the larger scales where our PAUS W3 jackknife would yield unreliable errorbars (see Sec. \ref{pau:sec:covariances}).
    }
    \label{pau:fig:pau_gama_wgp_wgg}
\end{figure*}

Projected correlation functions $w_{\rm{g}+}$, describing the radial alignment of galaxy shapes with galaxy positions, and $w_{\rm{gg}}$, the galaxy position auto-correlation, are displayed for PAUS W3 and GAMA in Fig. \ref{pau:fig:pau_gama_wgp_wgg}. For reasons detailed in Appendix \ref{pau:sec:app_zphrandoms} (where we also present successful systematics-tests $w_{\rm{g\times}}$), we elected to compare with GAMA the PAUS signals measured using our {\sc{LePhare}} colour-split (Fig. \ref{pau:fig:PAUGAMAzCMD}), our zph-windowed randoms (Sec. \ref{pau:sec:photoz_randomsimpact}), and the dynamic binning in line-of-sight separation $\Pi$ (Sec. \ref{pau:sec:clust_method}) -- the best-performing setup for our tests against the mock \pz GAMA sample (Sec. \ref{pau:sec:photoz_randomsimpact} \& Appendix \ref{pau:sec:app_zphrandoms}). The figure displays signals measured in PAUS W3 (solid pentagons/curves), for galaxies in the \pz range $0.1<z_{\rm{phot.}}<0.8$, along with spectroscopic signals from GAMA (open circles) and the signals measured in our mock \pz GAMA sample (downward triangles) using the same randoms/$\Pi$-binning setup for treatment of \pz errors. Grey-hatching indicates the regime where we would be unable to compute reliable errorbars for PAUS (see Sec. \ref{pau:sec:covariances}).

Even with our treatment, one sees the degradation of strongly significant red galaxy IA signals from spectroscopic GAMA data when measured using our mock photometric redshifts; we can expect the real, inaccessible PAUS IA signature to be similarly degraded by \pz scatter. Still, we are able to make a detection at $\sim3.1\sigma$ for red galaxy alignments in PAUS W3, and the significance of signals is consistently much higher than for blue galaxies (see Appendix \ref{pau:sec:app_zphrandoms}).

Given the results in the literature \citep{Mandelbaum06,Hirata2007,Joachimi2011,Johnston2019,Georgiou2019b}, one might expect to find even more significant radial alignments in this sample. \cite{Joachimi2011} found such correlations in photometric red galaxy samples of similar intrinsic brightness, over similar redshift ranges. Their samples, however, spanned a much larger area on-sky (more than $150\times$) with respect to PAUS W3, consequently yielding smaller statistical errors, and being comparatively dominated by bright central galaxies; \cite{Johnston2019} found central galaxy shapes to be the highest S/N contributor to alignment signals. However, they also found that satellite galaxies -- which will be more prevalent in PAUS -- do play a role in sourcing $w_{\rm{g+}}$ signals via their positions, particularly on smaller scales; we think it likely that such a small-scale signal may be found in PAUS with more area and continued development of \pz and their treatment.

To guide the eye, we computed some non-linear alignment model \citep[NLA;][]{Hirata2004a,Hirata2007,Bridle2007} predictions for $w_{\rm{g+}}$ using the PAUS-DEEP2 galaxies' spectroscopic $n(z)$, displaying these in Fig. \ref{pau:fig:pau_gama_wgp_wgg} as green/blue dashed lines (top-left panel) -- these curves are proportional to the product of a density sample galaxy bias $b_{\rm{g}}$ and an intrinsic alignment amplitude $A_{\rm{IA}}$. We find that the difference in NLA model predictions computed for $n(z)$ given by DEEP2 spec-$z$, or by {\sc{bcnz2}} photo-$z$, is negligible. For reference, the $A_{\rm{IA}}b_{\rm{g}}=6$ curve (green dashed) roughly corresponds to the constraints of \cite{Johnston2019}, obtained by fitting jointly to red-red galaxy clustering and red-red position-intrinsic shear correlations measured in GAMA data, with shapes from Kilo Degree Survey imaging \citep[KiDS;][]{Kuijken2019}.

%As might be expected, given the reasonable degree of similarity between the red samples in PAUS and GAMA indicated by Figs. \ref{pau:fig:PAUGAMAzCMD} \& \ref{pau:fig:gama_pau_magnitudes}, the $w_{\rm{g}+}$ signals agree rather well on the intermediate scales $1-20\mpc$, particularly when limiting PAUS to higher-quality photo-$z$, as in Fig. \ref{pau:fig:qz50signals}. Red GAMA galaxies, however, display a strong small-scale signal not seen in PAUS -- \cite{Johnston2019} found red satellite galaxies to be the principal drivers of this correlation (their Fig. 7). PAUS is deeper and has more intrinsically faint red objects than does GAMA (Fig. \ref{pau:fig:gama_pau_magnitudes}), so one might expect a large number of satellites and a significant signal. 

%The photo-$z$ scatter could be damping the signal -- one would indeed expect this effect to be stronger on small scales, and the artificially scattered GAMA galaxies do exhibit diluted signals, relative to the spectroscopic sample (large vs. small circles in Fig. \ref{pau:fig:pau_gama_wgp_wgg}). Whether our treatment of GAMA is insufficient to accurately mimic the loss of signal in PAUS, or whether the signal simply does not exist, are questions we leave to future work. A possible alternative explanation is that the mean redshift of red satellites in GAMA is significantly lower than for PAUS (Fig. \ref{pau:fig:PAUGAMAzCMD}) -- evolution of red satellite galaxies' small-scale alignments with redshift could result in such a discrepancy.

More interesting for the PAUS sample are the blue galaxy alignments, which are consistent with zero (for both surveys), at high precision. This result is especially interesting given the differences in redshift and luminosity distributions of blue galaxies in PAUS and GAMA (Figs. \ref{pau:fig:PAUGAMAzCMD} \& \ref{pau:fig:gama_pau_magnitudes}); we have yet more evidence for a lack of intrinsically aligned blue galaxies, on scales of $0.1-18\mpc$, regardless of their selection \citep{Hirata2007,Mandelbaum2011,Johnston2019}. Despite the small area of PAUS W3 ($\sim19\sqdeg$), the precision of blue galaxy signals is comparable with that from GAMA ($\sim180\sqdeg$), even after adding noise via our dynamic $\Pi$-binning, owing to the increased depth and consequently high number density of PAUS. It should be considered that the aforementioned \pz degradation of signals also applies here; however, the mock GAMA signals are noisier and less precisely null than the spectroscopic signals. Thus we might argue that the underlying signal for PAUS should be consistent with zero at high precision for the added noise to have so little effect. Indeed, every configuration we explore for the measurement of this blue galaxy IA signature yields signals that are non-zero at $\lesssim0.8\sigma$, typically not exceeding $\sim0.4\sigma$.

One might expect lower amplitudes of clustering in PAUS (compared with GAMA) for two reasons; (i) fainter galaxies are known to be less biased than brighter ones, and (ii) the photo-$z$ scatter will create a general suppression of power, particularly on smaller scales. Therefore, taken at face-value, the PAUS blue galaxy clustering signal in Fig. \ref{pau:fig:pau_gama_wgp_wgg} is of surprisingly high amplitude with respect to spectroscopic GAMA; blue galaxies in PAUS are on average much fainter than those in GAMA (Fig. \ref{pau:fig:gama_pau_magnitudes}), and subject to \pz errors which are expected to lower amplitudes and perhaps flatten power laws (by more efficiently suppressing small-scale signals). Complicating matters further, these data-points are highly correlated, making by-eye comparisons less instructive. We summarise our clustering results for each different correlation configuration in Appendix \ref{pau:sec:app_zphrandoms}, and state here our general findings: blue galaxy clustering in PAUS is seen to exhibit a shallower gradient from small to large scales, with respect to GAMA, and to have typically smaller or comparable amplitudes at $r_p=1\mpc$, except when using dynamic $\Pi$-bins \emph{and} selecting on \pz quality via \ttt{Qz} (see Fig. \ref{pau:fig:clust_summary_fig}) -- the use of either dynamic bins or \pz selection alone yields similar amplitudes/slopes, with little sensitivity to the choice of randoms.

As Fig. \ref{pau:fig:gama_pau_magnitudes} shows, red galaxies in PAUS are of roughly similar intrinsic brightness to those in GAMA, making the high amplitude of that signal less concerning. We see a steeper power law with respect to GAMA, going against the prediction for \pz degradation -- curiously, we see similarly steep correlations for the mock GAMA sample, perhaps signalling a differential impact of our mock \pz errors coming from magnitude-colour correlations inherent to any galaxy sample; redder galaxies tend to be brighter, such that Eqs. \ref{pau:eq:pauslike_photoz_scatter} \& \ref{pau:eq:pauslike_photoz_Pkick} will behave differently for red and blue objects. However, the aforementioned concerns with respect to correlations between $r_{p}$-scales also apply here, thus we neglect to investigate further, noting that a more realistic mock \pz implementation will aid with disentangling these results. We generally find greater consistency with spectroscopic GAMA clustering for red galaxies, with only slightly steeper gradients and slightly lower amplitudes -- see Appendix \ref{pau:sec:app_zphrandoms} for a more complete breakdown of signals from our various correlation configurations.

In future work, we will repeat and improve upon this analysis with the full PAUS area; the statistical power offered by PAUS will then be competitive with GAMA, providing new avenues and physical scales for directly constraining intrinsic alignment models -- such as the IA halo model with specific prescriptions for red and blue, central and satellite fractions, as they evolve with redshift \citep{Fortuna2020}. A mock \pz realisation, with higher fidelity to the error distribution exhibited by PAUS, will help us to refine our techniques for extraction of noisy IA signals from these data, and allow us to make stronger statements regarding the nature of alignments in the weakly non-linear regime.

\section{Discussion}
\label{pau:sec:discussion}

We have presented the first correlation functions from the Physics of the Accelerating Universe Survey, having measured the projected 3D intrinsic alignments and clustering of galaxies. These data are unique for their high-quality point-estimates of photometric redshifts, from 40-narrow-band optical photometry. The resulting W3-field catalogue allows for the deepest direct, flux-limited study of intrinsic alignments to date.

We calculated a unique set of $k$-corrections per object, by arbitrarily redshifting the best-fit SED models from our photo-$z$ pipeline. Scattering numerous clones of real galaxies uniformly within their respective $V_{\rm{max}}$ -- computed using the $k$-corrections and survey flux-limit -- we created random galaxy catalogues which reproduce the un-clustered radial selection function of the data. We also produced `windowed' randoms, wherein clones are scattered within a Gaussian window centred on the parent galaxy's redshift; in this way, the randoms mimic the preservation of galaxy populations across the redshift range, naturally encoding any luminosity-evolution.

Going further, we extended the formalism of $V_{\rm{max}}$-based randoms catalogue generation to encode the photometric redshift error distribution of PAUS, testing our methods against mock GAMA \pz samples. \tbf{By sampling redshifts from a probability density function described by the available spectroscopic $n(z)$ surrounding each galaxy's photo-$z$}, we generated probability density distributions for $V_{\rm{max}}$ from which to create ensemble randoms that compensate catastrophic failures in photometric redshift determination. We demonstrated that these \pz randoms aid in the recovery of the form of spectroscopic clustering signatures when measured in our mock \pz sample, though the benefits for intrinsic alignment signal recovery are limited. Our methods should scale well with increased spectroscopic sample densities for more accurate characterisation of the photometric redshift errors.

Splitting the PAUS W3 sample into red and blue galaxies, we measured the projected position-intrinsic shear and position-position correlations for $0.1<z_{\rm{phot.}}<0.8$, comparing our signals to analogous measurements from the spectroscopic GAMA survey and our mock \pz GAMA sample. On the intermediate scales $0.1<r_{p}<18\mpc$, \tbf{we made detections of red galaxy alignments at $\sim3\sigma$, though we think it likely that still stronger correlations are present but lost to photometric redshift degradation, similar to what we observe for the mock GAMA sample}. For blue galaxies, we find null signals at high precision, in support of the growing body of literature suggesting blue galaxies not to be aligned with the large-scale structure. Whilst the effects of \pz could be washing-out correlations, we argue that the consistently low noise-levels and significances exhibited by the signals are suggestive of an underlying null signal.

Relative to spectroscopic GAMA signals, we find red galaxies in PAUS W3 to cluster similarly, but for slightly lower amplitudes and steeper clustering gradients. Our use of randoms that account for \pz errors strengthens consistency with GAMA, and the slope of observed clustering power laws is sensitive to our choices with respect to binning of signals over the line-of-sight, prior to projection. For blue galaxies in PAUS W3, with find consistently shallower clustering gradients with respect to GAMA, possibly resulting from \pz effects -- this flattening of gradients is mirrored by our mock \pz GAMA signals -- and a similar sensitivity of amplitudes to our line-of-sight binning choices.

%Selecting the best half of photo-$z$ in PAUS W3, we re-measured all signals to find a remarkable level of agreement between PAUS and GAMA; the consistency of IA signals is slightly strengthened, and the PAUS clustering amplitudes rise to match those from GAMA extremely closely -- this result is interesting, given that whilst the samples are similar in intrinsic luminosity, PAUS does have significantly more faint objects.

Future work will feature: greatly increased statistical power, courtesy of more than double the current area; a more detailed assessment of the photo-$z$ suppression of correlations in PAUS; fits of halo/other models to the observed clustering and IA correlations; and an assessment of luminosity/redshift-dependence in the signals, making use of our windowed-clone, photo-$z$-compensating random galaxy catalogues -- all of this will add to the growing understanding of IA within the weak lensing community, informing the choices of modelling for the next generation of cosmic shear analyses.

{\small
\textbf{Acknowledgements:} we acknowledge helpful discussions with Danny Farrow and members of the PAUS collaboration. Authors are listed in 3 groups, the last 2 alphabetically: (i) the main authors of this work, (ii) team members who have significantly contributed to the production of data products used in this paper and/or have provided substantial feedback on the paper, and (iii) builders of the PAU Survey and tertiary contributors.

The authors are grateful to the developers of \galsim\ for making their software packages publicly available. HJ acknowledges support from a UK STFC studentship, and from the Delta ITP consortium, a programme of the Netherlands Organisation for Scientific Research (NWO) that is funded by the Dutch Ministry of Education, Culture and Science (OCW). PN acknowledges support from ST/P000541/1 and ST/T000244/1. HHo acknowledges support from  the Netherlands Organisation for Scientific Research (NWO) through grant 639.043.512. ME, JA acknowledge funding from the European Union’s Horizon 2020 research and innovation programme under grant agreement No 776247. HHi is supported by a Heisenberg grant of the Deutsche Forschungsgemeinschaft (Hi 1495/5-1) as well as an ERC Consolidator Grant (No. 770935). MS acknowledges funding from the European Union's  Horizon 2020 research and innovation programme under the Maria Skłodowska-Curie (grant agreement No 754510), the National Science Centre of Poland (grant UMO-2016/23/N/ST9/02963) and by the Spanish Ministry of Science and Innovation through Juan de la Cierva-formacion programme ( reference FJC2018-038792-I). GM acknowledges support from ST/P006744/1.

The PAU Survey is partially supported by MINECO under grants CSD2007-00060, AYA2015-71825, ESP2017-89838, PGC2018-094773, SEV-2016-0588, SEV-2016-0597, and MDM-2015-0509, some of which include ERDF funds from the European Union. IEEC and IFAE are partially funded by the CERCA programme of the Generalitat de Catalunya. Funding for PAUS has also been provided by Durham University (via the ERC StG DEGAS-259586), ETH Zurich, Leiden University (via ERC StG ADULT-279396 and Netherlands Organisation for Scientific Research (NWO) Vici grant 639.043.512), University College London and from the European Union's Horizon 2020 research and innovation programme under the grant agreement No 776247 EWC. The PAU data centre is hosted by the Port d'Informaci\'o Cient\'ifica (PIC), maintained through a collaboration of CIEMAT and IFAE, with additional support from Universitat Aut\`onoma de Barcelona and ERDF. We acknowledge the PIC services department team for their support and fruitful discussions.

GAMA is a joint European-Australasian project based around a spectroscopic campaign using the Anglo-Australian Telescope. The GAMA input catalogue is based on data taken from the Sloan Digital Sky Survey and the UKIRT Infrared Deep Sky Survey. Complementary imaging of the GAMA regions is being obtained by a number of independent survey programmes including GALEX MIS, VST KiDS, VISTA VIKING, WISE, Herschel-ATLAS, GMRT and ASKAP providing UV to radio coverage. GAMA is funded by the STFC (UK), the ARC (Australia), the AAO, and the participating institutions. The GAMA website is http://www.gama-survey.org/ . Based on observations made with ESO Telescopes at the La Silla Paranal Observatory under programme ID 177.A-3016.
}

\bibliographystyle{aa}
\bibliography{PAUSrefs.bib}{}

\appendix

\section{`zph-randoms' \& $\Pi$-binning}
\label{pau:sec:app_zphrandoms}

\begin{figure*}[!htp]
    \centering
    \includegraphics[width=\textwidth]{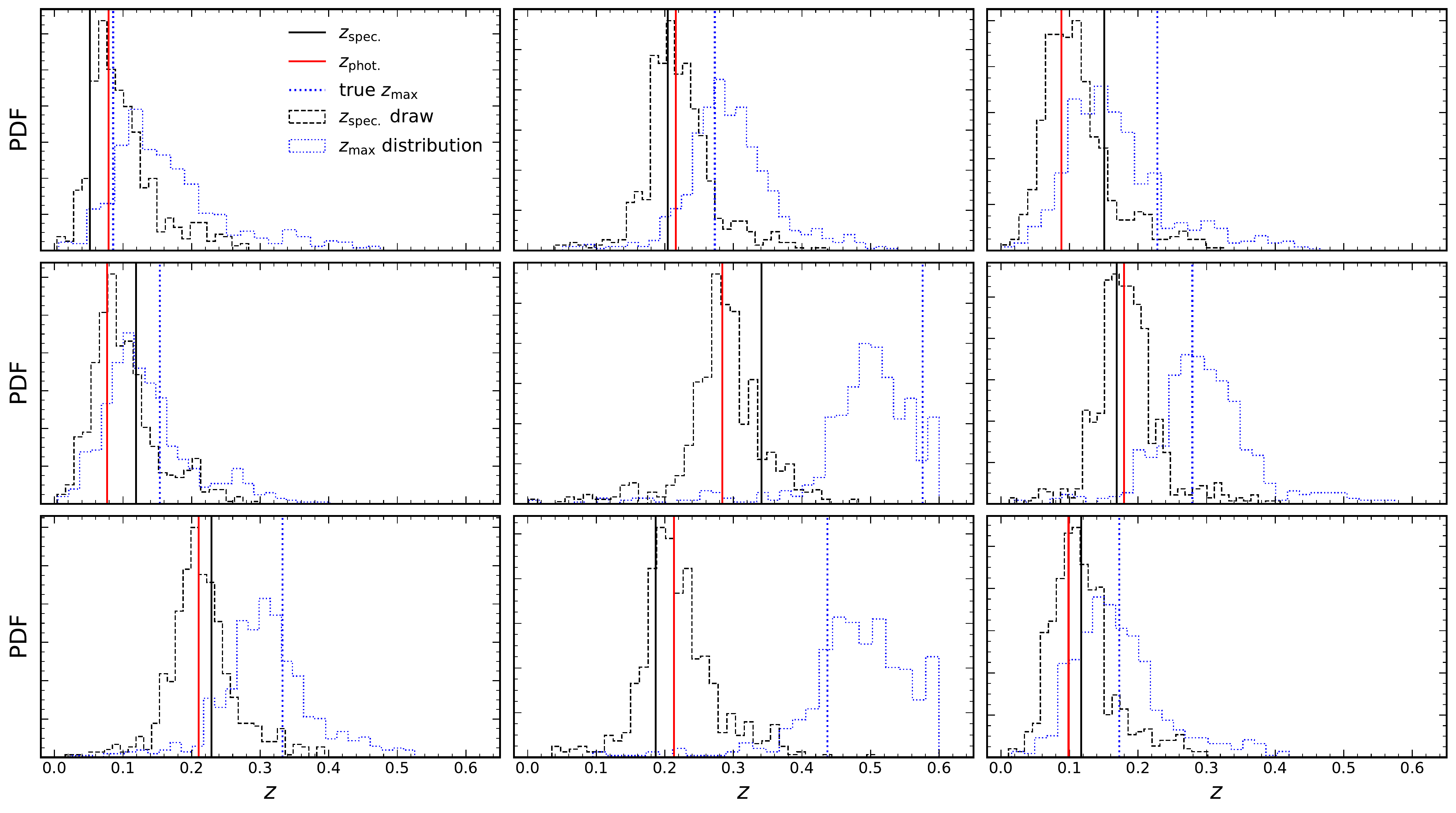}
    \caption{Artificial photometric (red vertical lines; Sec. \ref{pau:sec:photoz_randomsimpact}), external spectroscopic (black vertical lines) and spectroscopically-calculated maximum (blue dotted vertical lines; Sec. \ref{pau:sec:vmax_randoms}) redshifts of 9 galaxies in GAMA. The realisations of GAMA spec-$z$, drawn from the conditional $n(z_{\rm{spec.}}\,|\,z_{\rm{phot.}}\pm0.03)$ distribution surrounding each galaxy's $z_{\rm{phot.}}$, are displayed as black dashed histograms, and the corresponding $z_{\rm{max}}$ distribution is given in each panel as a blue dotted histogram. One sees that errors in \pz (red vs. black vertical lines) and the corresponding inferred $z_{\rm{max}}$ (peaks of blue histograms vs. blue vertical lines) are compensated by the draws from $n(z_{\rm{spec.}}\,|\,z_{\rm{phot.}})$ and resulting $z_{\rm{max}}$ distributions.
    }
    \label{pau:fig:zspeczmax_draws}
\end{figure*}

%\begin{figure}[!htpb]
%    \centering
%    \includegraphics[width=\columnwidth]{signalratios_gama_zphot_2_mixed.pdf}
%    \caption{The ratio of intrinsic alignment (\emph{top}; for red galaxies) and clustering (\emph{bottom}; for all galaxies) signals, measured in our mock \pz GAMA galaxy sample under various randoms/$\Pi$-binning scenarios (Secs. \ref{pau:sec:photoz_randomsimpact} \& \ref{pau:sec:clust_method}), to the equivalent spectroscopically-measured signals, $1\sigma$ jackknife errors for which \citep[see][]{Johnston2019} are indicated by shading. One sees that the dynamic binning in $\Pi$ (circles) recovers much of the signal lost due to \pz scatter (triangles) in both panels, and that the use of zph-randoms (dot-dashed) and windowing (filled points) further aid in recovering the shape of the clustering signal.
%    }
%    \label{pau:fig:signalratios_zphrandoms}
%\end{figure}

\begin{sidewaysfigure*}[!htpb]
    \centering
    \includegraphics[width=\textwidth]{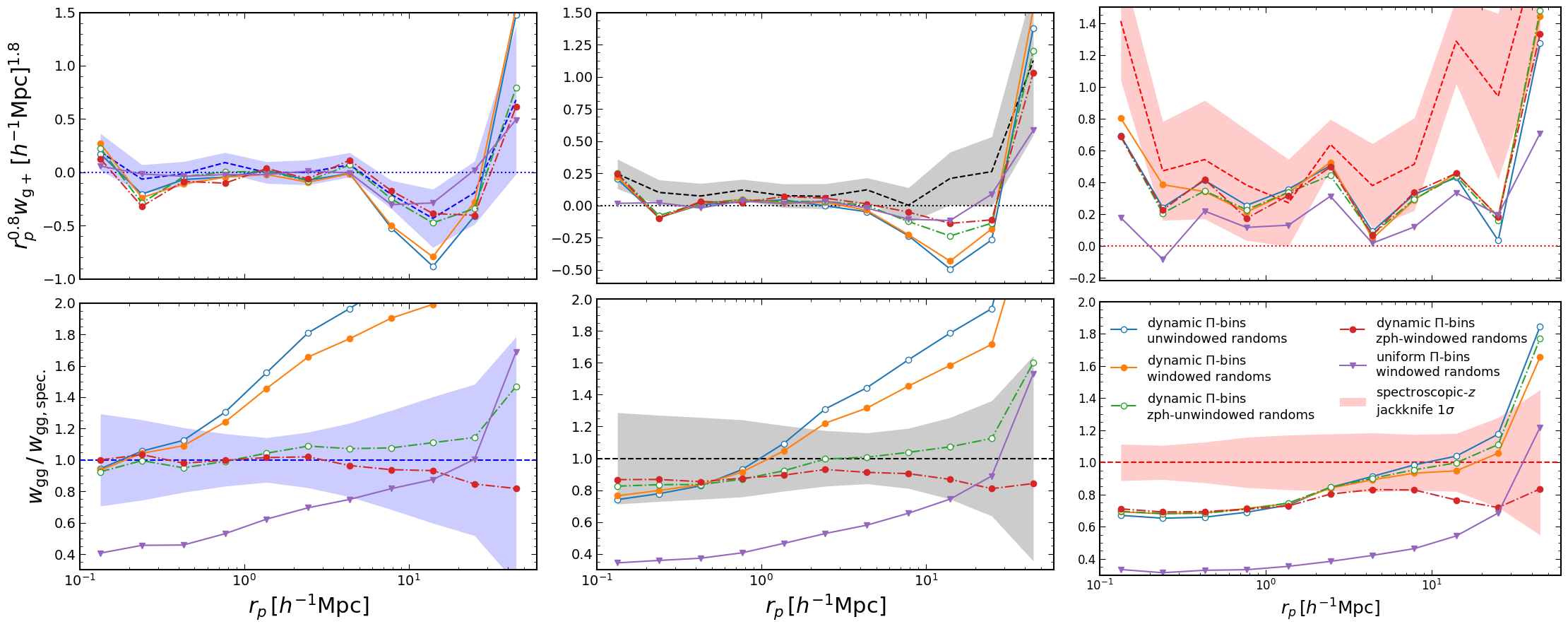}
    \caption{Attempted recoveries of all red (\emph{right}), blue (\emph{left}) and total (\emph{middle}) sample correlations measured in our mock \pz GAMA galaxy sample under various randoms and $\Pi$-binning scenarios (Secs. \ref{pau:sec:photoz_randomsimpact} \& \ref{pau:sec:clust_method}), with IA signals compared directly (\emph{top}), and clustering signals shown in ratio (\emph{bottom}), to the equivalent spectroscopic signals (dashed lines). $1\sigma$ jackknife errors \citep[see][]{Johnston2019} are indicated by shading. One sees that the dynamic binning in $\Pi$ (circles) recovers much of the non-zero IA and clustering signal lost due to \pz scatter (triangles), and that the use of zph-randoms (dot-dashed) and windowing (filled points) further aids in recovering the shape of the clustering signal, concerns with respect to amplitudes notwithstanding.
    }
    \label{pau:fig:signalbreakdown_zphrandoms}
\end{sidewaysfigure*}

\begin{figure*}[!htpb]
    \centering
    \includegraphics[width=\linewidth]{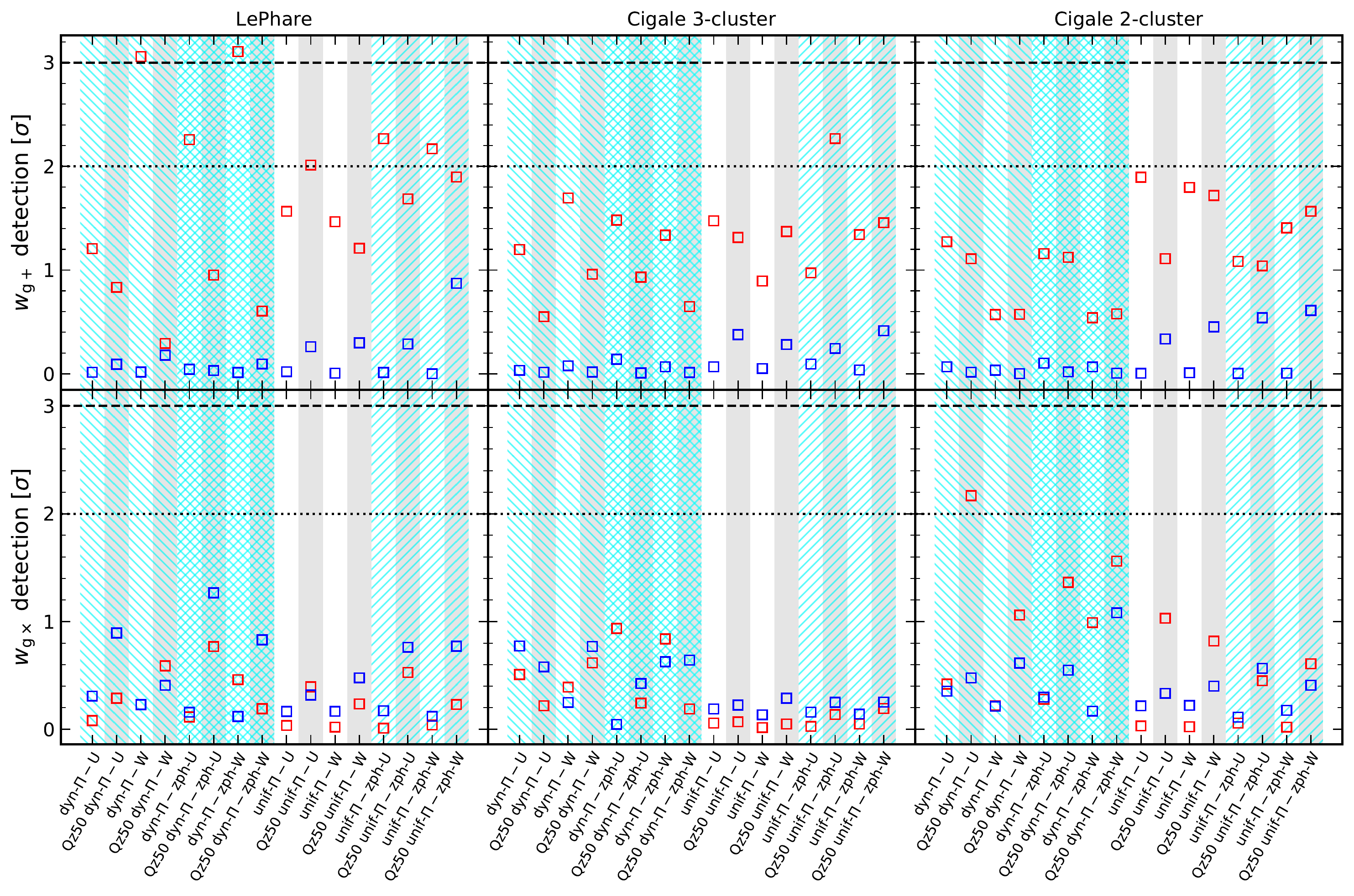}
    \caption{Summary of the significance of projected intrinsic alignment signal detection (given as $x\sigma$) in PAUS W3, with radial alignment component $w_{\rm{g+}}$ on the top row and cross-component (systematics-test) $w_{\rm{g\times}}$ on the bottom. Significances are computed in the range $0.1<r_{p}<18\mpc$ using full jackknife covariances (Sec. \ref{pau:sec:covariances}). Red and blue galaxy correlations are denoted by marker colours. The 3 columns show results for our different schemes of sample colour-splitting. Labels on the x-axis give the remaining specifics of each correlation configuration; `U' (`W') denotes unwindowed (windowed) randoms. For ease of comparison, cyan forward-slash hatching indicates setups using zph-randoms (Sec. \ref{pau:sec:photoz_randomsimpact}), and back-slashes indicate dynamic $\Pi$-binning (as opposed to uniform binning; no back-slashes). Grey shading denotes correlations measured on the best 50\% of galaxies by \pz quality (selected on the \ttt{Qz} parameter). We mark confidence levels $2\sigma$ and $3\sigma$ with horizontal lines. One sees that several correlation configurations yield red galaxy alignment signals reaching $\sim2-3\sigma$ in significance. We make no significant detections of blue galaxy alignments, or systematics-test correlations, in the PAUS W3 data.
    }
    \label{pau:fig:IA_summary_fig}
\end{figure*}

\begin{figure*}[!htpb]
    \centering
    \includegraphics[width=\linewidth]{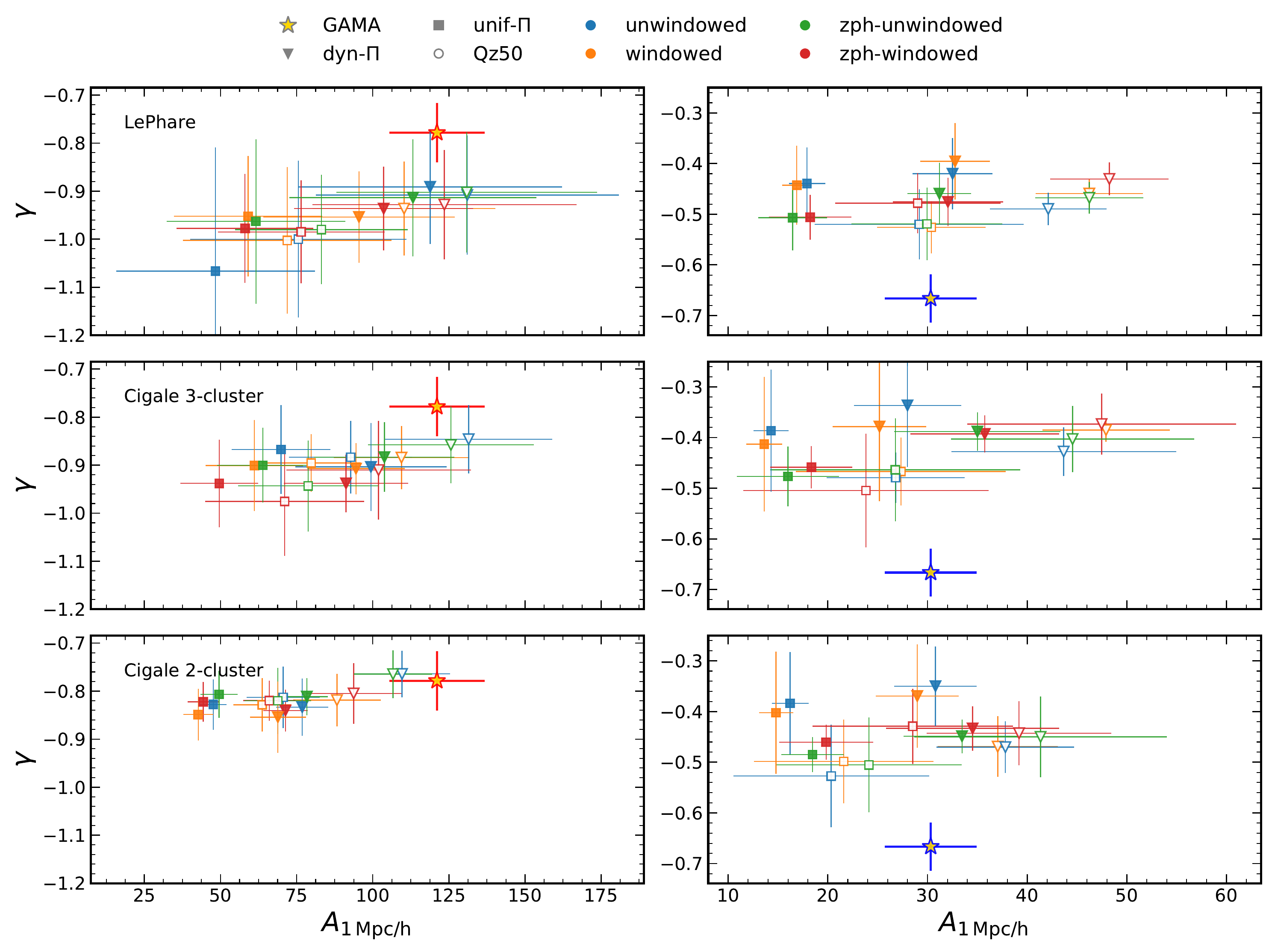}
    \caption{Comparison of projected clustering signal measurements in PAUS W3 and in spectroscopic GAMA data, summarised as fits of the simple, 2-parameter (amplitude $A$ at $r_p=1\mpc$, and slope $\gamma$) power law given by Eq. \ref{pau:eq:wgg_powerlaw}. Fits are made via the non-linear least squares method, in the range $0.1<r_{p}<18\mpc$, and using full jackknife covariances (Sec. \ref{pau:sec:covariances}). All measured clustering signals feature heavily correlated $r_{p}$-scales, sometimes resulting in poor goodness-of-fit, thus these points should be taken with moderation. Red (blue) galaxy correlations are shown in the left-hand (right-hand) column, with GAMA correlations given by gold stars with red (blue) outlines. PAUS correlations from each of our various configurations (changing $\Pi$-binning, randoms and \pz selection) are denoted by markers and colours, as indicated by the legend.
    }
    \label{pau:fig:clust_summary_fig}
\end{figure*}

\begin{figure*}[!htbp]
    \centering
    \includegraphics[width=\linewidth]{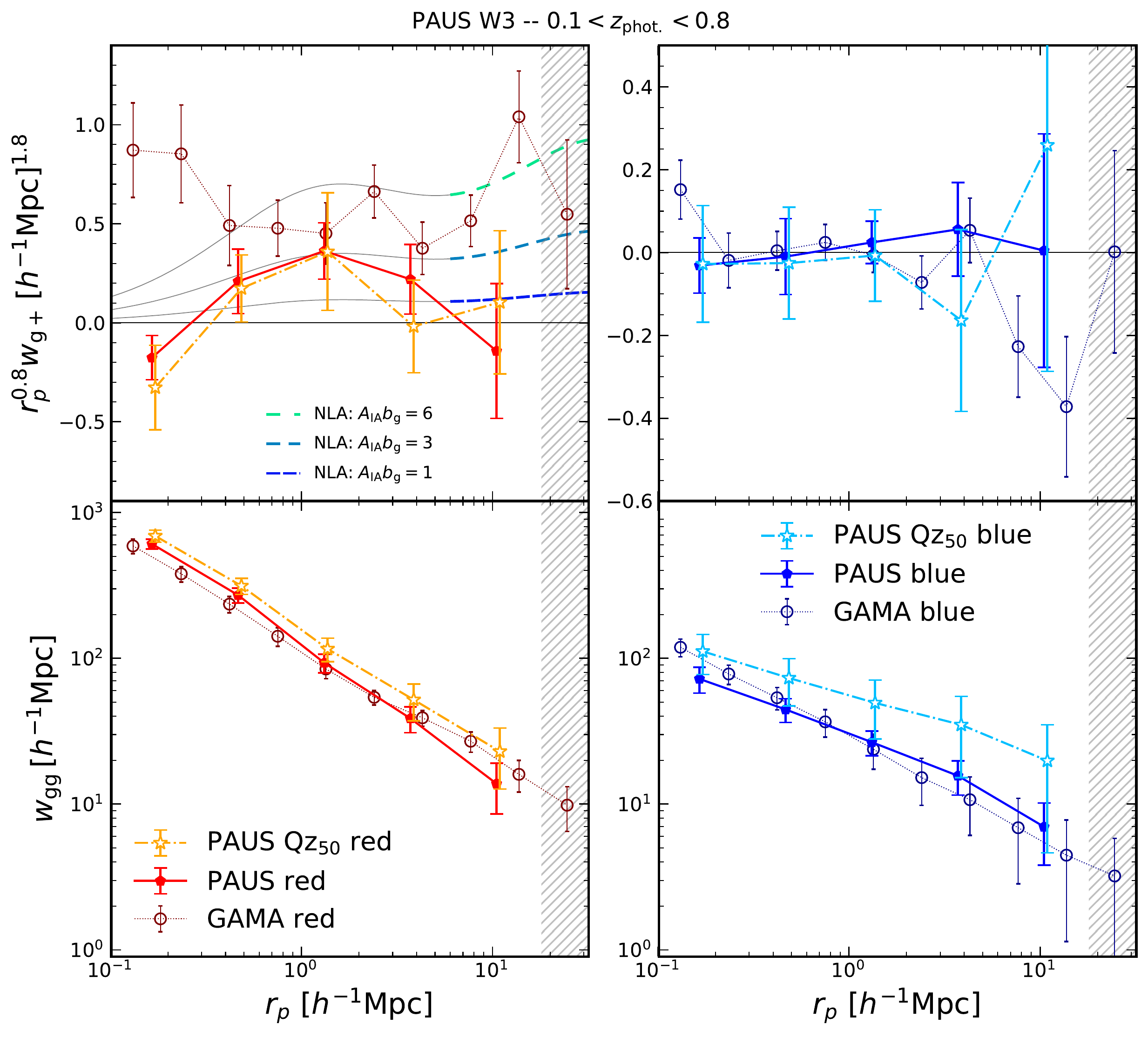}
    \caption[PAUS and GAMA: projected position-intrinsic shear and position-position correlations for red/blue galaxies in PAUS (selected for \pz quality) and GAMA]{Same as Fig. \ref{pau:fig:pau_gama_wgp_wgg}, but swapping the mock \pz GAMA signals out for a comparison with signals from the best 50\% of PAUS \pz according to the \ttt{Qz} parameter (denoted by `Qz$_{50}$' in the legends, and open-star data-points). Increases in noise and clustering amplitudes are evident when selecting on \pz quality in this way.
    }
    \label{pau:fig:pau_gama_wgp_wgg_qz50}
\end{figure*}

Here we include some additional discussion and figures detailing the construction and testing of the `zph-randoms' we introduced in Sec. \ref{pau:sec:photoz_randomsimpact}, and also the dynamic $\Pi$-binning from Sec. \ref{pau:sec:clust_method}. Our objective here was to recover as well as possible the form of the spectroscopic IA and clustering signals from GAMA, using our approximately PAUS-like mock photo-$z$ (see Sec. \ref{pau:sec:photoz_randomsimpact}). 

Fig. \ref{pau:fig:zspeczmax_draws} displays, for a random selection of 9 GAMA galaxies, our artificial, PAUS-like $z_{\rm{phot.}}$ (red vertical lines; see Sec. \ref{pau:sec:photoz_randomsimpact}), the galaxy $z_{\rm{spec.}}$ from GAMA (black vertical lines), and the 320 realisations of `spectroscopic' redshifts that we draw for each galaxy from conditional distributions $n(z_{\rm{spec.}}\,|\,z_{\rm{phot.}}\pm0.03)$, shown as black dashed histograms. One sees that redshifts closer to the parent $z_{\rm{spec.}}$ are drawn in each case. The resulting distributions of $z_{\rm{max}}$ (Eq. \ref{pau:eq:zmax_relation}) are given as blue dotted histograms, the peaks of which correspond to the starting $z_{\rm{phot.}}$ -- comparing those peaks/distributions to the true $z_{\rm{max}}$ (blue vertical lines) calculated from $z_{\rm{spec.}}$, one sees that photo-$z$-induced errors in $z_{\rm{max}}$-determination are compensated by our procedure here. The resulting, ensemble zph-randoms that we created are thus smoother in their final $n(z)$ (see Fig. \ref{pau:fig:randoms}), correcting obvious excesses at low redshifts and capturing our hand-made redshift degeneracies (Fig. \ref{pau:fig:nzszp_marginals}).

The utility of zph-randoms is most clearly demonstrated in Fig. \ref{pau:fig:signalbreakdown_zphrandoms}, where we compare the IA and clustering signals measured with our artificial GAMA \pz to those measured with spectroscopic redshifts. The top-panels show the IA signals $w_{\rm{g+}}$, with the clustering $w_{\rm{gg}}$ given in the bottom-panels. Each clustering signal is shown in ratio to the spectroscopic signal, whilst IA signals are simply overlain due to their low signal-to-noise. The $1\sigma$ jackknife errors \citep[see][for details]{Johnston2019} for the spectroscopic signals are given by the shaded regions. The various scenarios of uniform or dynamic $\Pi$-binning (Sec. \ref{pau:sec:clust_method}), and standard or zph-randoms, are indicated in the legend. One sees that both the binning and the choice of randoms are highly consequential for the clustering measurements, with the dynamic $\Pi$-bins and zph-windowed randoms (red circles) best recovering the shape of the clustering correlation function, and its amplitude to well within the spectroscopic error (at least for the total and blue samples) -- residual disagreements in amplitude should be easier to accommodate with \pz modelling, compared with large differences in shape arising due to redshift outliers. The IA signal is clearly more difficult to recover with fidelity, likely due to the characteristically low signal-to-noise of these correlations and their easy erasure by photo-$z$ scatter.

Having tested our \pz treatment with mock GAMA samples, we had to decide which combination of methods to apply to measurements of equivalent signals in PAUS. We computed the projected IA and clustering signals for every permutation of the following choices:

\begin{itemize}
    \item Colour-split: samples defined according to {\sc{LePhare, Cigale}} 2-cluster, or {\sc{Cigale}} 3-cluster (Fig. \ref{pau:fig:PAUGAMAzCMD});
    \item $\Pi$-binning: `standard', uniform bins, or dynamic (adapted Fibonacci sequence; Sec. \ref{pau:sec:clust_method}) bins;
    \item Randoms: unwindowed, windowed, zph-unwindowed, or zph-windowed;
    \item Photo-$z$ selection: best 50\% on \ttt{Qz} parameter (Fig. \ref{pau:fig:zphot_zspec_qz}), or no selection,
\end{itemize}
for a total of 48 configurations. Displaying all of these, often similar, correlations poses difficulties for meaningful comparison, hence we elect to summarise the statistics in the following figures. 

For projected IA signals $w_{\rm{g+}}(r_{p})$ [and systematics-tests $w_{\rm{g\times}}(r_{p})\,$], we display the significance of non-zero detection (computed with the full jackknife covariance\footnote{For this calculation, and the clustering fits to be described, we must apply a correction factor \citep[see][]{Hartlap11} to noisy estimates of the inverse covariance, given as $(N - D - 2) / (N - 1)$, where $N$ is the number of jacknife samples (24) and $D$ is the number of data-points (5); in the limit $N\gg{}D$, the correction approaches unity.}) in the range $0.1<r_{p}<18\mpc$, where we are able to compute reliable jackknife errors for the PAUS W3 area -- see Sec. \ref{pau:sec:covariances}. This IA summary is displayed in Fig. \ref{pau:fig:IA_summary_fig}, \tbf{where one sees that we make no significant detection of systematics indicated by the position-shear correlation cross-component $w_{\rm{g\times}}$, and that several configurations yield red galaxy alignment signals exceeding $2\sigma$ in signifance, with the most significant detection at $3.1\sigma$}. The three columns in this figure indicate the colours used to define red and blue galaxy samples, as given by the column-title. The top row gives $w_{\rm{g+}}$ correlations, with $w_{\rm{g\times}}$ systematics-tests on the bottom. Different correlation configurations are indicated at the bottom of the figure, with reference to $\Pi$-binning choices (`dyn'-amic, or `unif'-orm), randoms choices (Unwindowed, Windowed, zph-Unwindowed, or zph-Windowed), and \pz selection (selected on \ttt{Qz}, or not).

\tbf{One sees generally stronger $w_{\rm{g+}}$ correlations for red galaxies defined with the {\sc{LePhare}} colour-split (see Fig. \ref{pau:fig:PAUGAMAzCMD}), though all red $w_{\rm{g+}}$ signals exceed their blue counterparts in significance.} Importantly, the significance of any $w_{\rm{g+}}$ detection for blue galaxies is always negligible, at $\lesssim0.8\sigma$ in all cases; modulo the impact of \pz signal suppression (which will be elucidated in a future analysis using all of the PAUS area), we have reaffirmed the literature findings of negligible alignments between blue galaxies and the galaxy density field, now for a fainter flux-limited sample of blue galaxies than has previously been considered \citep[see][]{Mandelbaum2011,Tonegawa2017,Johnston2019}.

For projected galaxy clustering signals, a meaningful comparison between correlation configurations is tricky, as we always expect significant detections due to the nature of the clustered galaxy distribution. Here, we perform non-linear least squares fits of simple, 2-parameter power laws of the form

\begin{equation}
    w_{\rm{gg}} = A\,\left(\frac{r_{p}}{1\mpc}\right)^{\gamma} \quad ,
    \label{pau:eq:wgg_powerlaw}
\end{equation}
to the clustering signals, again in the range $0.1<r_{p}<18\mpc$. The free parameters in the fit are the amplitude $A$ at $r_{p}=1\mpc$, and the power law slope $\gamma$, and we again make use of the full jackknife covariance for the fit. We display the fitted parameters for each of our 48 correlation configurations, and analogous fits (same scale-cuts) made to the spectroscopic GAMA signals, in Fig. \ref{pau:fig:clust_summary_fig}. Rows here give the different colour-split choices, with columns separating red (left) and blue (right) clustering fits. The remaining choices ($\Pi$-binning, randoms, \pz selection) are indicated by markers and colours, as detailed in the legend, and GAMA fits are shown as gold stars with red or blue edges. In general, we see steeper power laws for red galaxy clustering, with respect to GAMA, and typically lower amplitudes at $r_{p}=1\mpc$. Conversely, we find flatter power laws for blue galaxies, and that the fitted amplitudes are sensitive to our choices of $\Pi$-binning (dynamic bins $\rightarrow$ larger amplitude) and \pz selection (best 50\% \ttt{Qz} $\rightarrow$ larger amplitude).

We elected to display, in Fig. \ref{pau:fig:pau_gama_wgp_wgg}, PAUS signals split by {\sc{LePhare}} colours, for the dynamic $\Pi$-bins and zph-windowed randoms setup favoured in testing with the mock GAMA sample. For clustering in particular, these signals should be closer to the inaccessible (spectroscopic) truth, provided that (i) our mock \pz are sufficiently PAUS-like, and (ii) the differences between PAUS/GAMA sample characteristics (area, depth, galaxy properties; see also Fig. \ref{pau:fig:gama_pau_magnitudes}) are not too impactful. Potential weaknesses in these assumptions include that (i) PAUS \pz performance correlates somewhat with galaxy type, which we do not attempt to account for, and that (ii) the survey targets fainter objects at higher redshifts. We shall explore the robustness of our assumptions here in future work.

We choose to display the {\sc{LePhare}} correlations due to the seemingly more discriminatory -- and GAMA-esque -- colour distribution inferred by {\sc{LePhare}}, as compared with {\sc{Cigale}} (Fig. \ref{pau:fig:PAUGAMAzCMD}). Indeed, the higher significance of $w_{\rm{g+}}$ signals measured in red galaxy samples thus defined (Fig. \ref{pau:fig:IA_summary_fig}) is encouraging; from the intrinsic alignments literature, we have very good reason to expect a radial alignment of these galaxies with the density field \citep{Mandelbaum2006,Hirata2007,Joachimi2011,Samuroff2018,Johnston2019,Georgiou2019b}. We reason then that our {\sc{Cigale}} colour classification scheme is not yet able to identify red galaxies as efficiently as the simple cut defined upon the {\sc{LePhare}} colour-magnitude plane (Fig. \ref{pau:fig:PAUGAMAzCMD}). Forthcoming work will investigate further, and present finalised analyses of PAUS with {\sc{Cigale}} (Siudek et al. in prep.).

\tbf{With our chosen setup, we made our most significant detection of red galaxy alignments in PAUS W3, at $3.1\sigma$. The many tentative detections shown in Fig. \ref{pau:fig:IA_summary_fig}, and the suppression of IA in our mock \pz signals (Fig. \ref{pau:fig:pau_gama_wgp_wgg}; top-left, open circles vs. downward triangles), suggest that a stronger signal may be present and washed-out by photo-$z$ errors.} We display \ttt{Qz}-selected sample correlations in Fig. \ref{pau:fig:pau_gama_wgp_wgg_qz50} for the same randoms/binning setup as in Fig. \ref{pau:fig:pau_gama_wgp_wgg}, noting the poor signal-to-noise for IA measurements. Moreover, the blue galaxies in PAUS are intrinsically fainter than in GAMA (Fig. \ref{pau:fig:gama_pau_magnitudes}), such that the higher observed amplitude of clustering is unlikely. It may be that our dynamic $\Pi$-binning is unsuitable for these more secure redshifts, though more investigation is required. A forthcoming analysis will extend our methods to the full PAUS area, making use of the latest \pz and exploring the potential for more accurate recovery of photo-$z$-suppressed two-point correlations.

\end{document}